\documentclass[letterpaper,aps,prl,reprint,citeautoscript,twocolumn,groupedaddress,superscriptaddress,longbibliography]{revtex4-2}

\usepackage[english]{babel}
\usepackage{amsmath,amssymb}
\usepackage{multirow}
\usepackage{bm}
\usepackage{mathtools}
\usepackage{dsfont}
\usepackage{amsfonts}
\usepackage[normalem]{ulem}
\usepackage{url}
\usepackage{wasysym}
\usepackage{bbm}
\usepackage{quantikz}
\usepackage{float} 
\usepackage{bibunits} 
\usepackage{MnSymbol}
\usepackage{times}
\usepackage{etoolbox}
\makeatletter
\renewcommand{\@bibunitname}{\jobname-bu\the\@bibunitauxcnt}
\AtBeginDocument{%
    \patchcmd{\bibliography}{\immediate\openout\@bibunitaux bu.aux}%
        {\immediate\openout\@bibunitaux\jobname-bu.aux}{}%
        {\PackageError{Style companion}{Cannot isolate bibliography output}{}}%
    \patchcmd{\bibliography}{\string\@input{bu.aux}}%
        {\string\@input{\jobname-bu.aux}}{}%
        {\PackageError{Style companion}{Cannot isolate bibliography input}{}}%
    \let\std@bibliography\bibliography
}
\patchcmd{\endbibunit}{\@input{bu.aux}}{\@input{\jobname-bu.aux}}{}%
    {\PackageError{Style companion}{Cannot isolate supplemental references}{}}
\makeatother

\definecolor{tum}{HTML}{0065bd}
\definecolor{arxivcite}{HTML}{004D6B}
\definecolor{arxivlink}{HTML}{8c1515}

\usepackage[colorlinks=true,linkcolor=arxivlink,citecolor=arxivcite,urlcolor=arxivlink,hypertexnames=false]{hyperref}
\usepackage{orcidlink}

\AtBeginDocument{%
    
    \renewcommand{\addcontentsline}[3]{}%
}

\def\ba#1\ea{\begin{align}#1\end{align}}
\def\bg#1\eg{\begin{gather}#1\end{gather}}
\def\bpm{\begin{pmatrix}}
\def\epm{\end{pmatrix}}

\mathchardef\mhyphen="2D

\newcommand{\ourtitle}{Non-Equilibrium Dynamics of Fractionalization in a Fractal Spin Liquid}

\allowdisplaybreaks

\begin{document}
\title{\textbf{\ourtitle}}

\author{Shi Feng\,\orcidlink{0000-0002-0184-7049}}
\email{shi.feng@tum.de}
\affiliation{Technical University of Munich, TUM School of Natural Sciences, Physics Department, 85748 Garching, Germany}
\affiliation{Munich Center for Quantum Science and Technology (MCQST), Schellingstr. 4, 80799 M{\"u}nchen, Germany}
\author{Johannes Knolle\,\orcidlink{0000-0002-0956-2419}}
\affiliation{Technical University of Munich, TUM School of Natural Sciences, Physics Department, 85748 Garching, Germany}
\affiliation{Munich Center for Quantum Science and Technology (MCQST), Schellingstr. 4, 80799 M{\"u}nchen, Germany}
\affiliation{Blackett Laboratory, Imperial College London, London SW7 2AZ, United Kingdom}
\author{Willian M. H. Natori\,\orcidlink{0000-0002-0740-2956}}
\affiliation{Gleb Wataghin Institute of Physics, University of Campinas, Campinas,
São Paulo 13083-950, Brazil}


\begin{abstract}
Nonstandard geometry reshapes both the nature of fractionalized quantum matter and how it can be observed. We show that the Kitaev model on a dual Sierpinski gasket supports a gapless chiral spin liquid, and its fractionalization and topology can be revealed via quench dynamics starting from simple product states.  Destructive interference on the fractal lattice prevents flux-bound zero modes' hybridization, preserving the topology of fractionalized Majorana fermions despite initial states' extensive self-generated gauge-flux disorder. Products of chiral three-spin states can therefore access the chiral topology without resolving many of the fluxes or preparing the many-body ground state. The fractal geometry and the initial states' definite total spin parity yield a closed parton representation of corner-spin correlations, revealing sharp chiral edge dynamics and the long-time memory from Majorana corner modes. These complementary signatures connect the topology of fractionalization in fractal spin liquids to experimentally accessible spin dynamics from trivial initial states.
\end{abstract}

\maketitle

\emph{ \color{tum}{Introduction.--}}
Lattice geometry can shape both the phases of quantum spin liquids and the dynamics of their fractionalized excitations \cite{Lenggenhager2025,Dusel2025,GrushinRepellin2023,CasselaOrnellasHodson2023,Agarwala24}. Kitaev's exact mapping of spins to Majorana fermions and $\mathbb Z_2$ gauge fluxes \cite{Kitaev2006} makes this interplay accessible beyond the original honeycomb lattice. Extensions to decorated honeycomb \cite{YaoKivelson2007,OBrien2016} reveal how connectivity changes chirality and the Majorana spectrum. More recently, amorphous and polycrystalline networks have been shown to support spontaneous chiral spin liquids \cite{CasselaOrnellasHodson2023,GrushinRepellin2023}, while hyperbolic Kitaev models realize an unexpected gapless chiral spin liquid in negatively curved space \cite{Lenggenhager2025,Dusel2025}. These establish geometry as an important ingredient in the nature of fractionalized quantum matter.

Fractals offer a largely uncharted setting for spin liquids, with a hierarchy of internal boundaries across length scales \cite{ben2000,Bunde1991, Nakayama2003,tosatti2012fractals}. Such structure can generate localized and propagating excitations on the same graph without translation symmetry \cite{Pai2019,Agarwala2019,Osseweijer2024,Eek25}, including the extreme localization that leads to compact eigenstates \cite{Nandy2020,Hanafi2022}.
Apart from their theoretical novelty, recently realized synthetic fractal lattices make them experimentally promising \cite{Biesenthal2022,LiFractalFloquet2023,Grass2025,verstraten2025}. 
While the existence of new phases of fractal quantum systems have recently been discussed in different contexts \cite{Itamar85,Pai2019,Cenke22,Sarangi2024,Eek25}, whether the fractal geometry has any non-trivial effects on spin liquids or advantage in probing their fractionalization using synthetic quantum simulation \cite{Semeghini21,Ebadi2021,Ransford2026,Haghshenas2026,Pollmann26}, remains largely unexplored. 

\begin{figure}[t]
    \centering
    \includegraphics[width=0.9\linewidth]{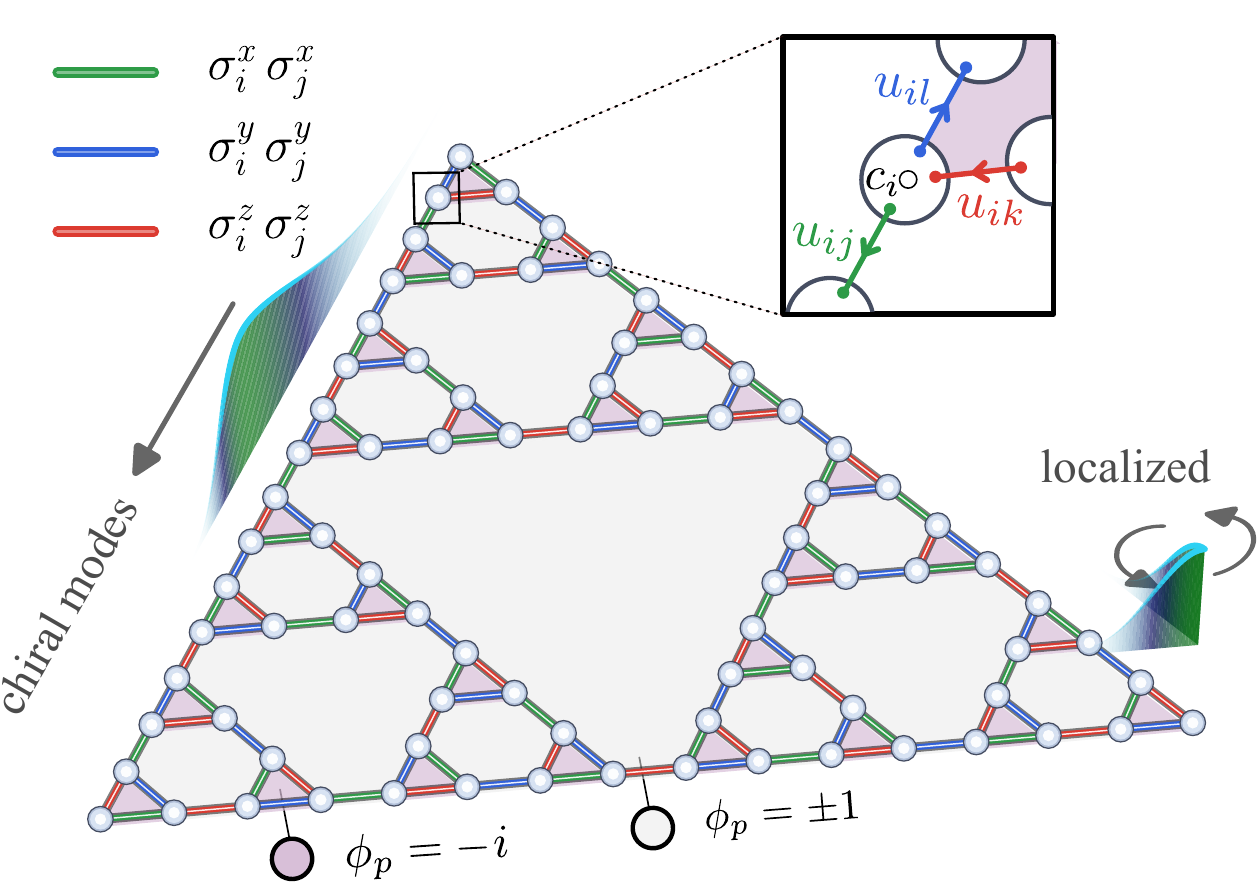}
    \caption{\textbf{Illustration of Kitaev-Sierpinski gasket.} Localized compact Majorana zero mode and propagating chiral modes coexist when even plaquettes have random unresolved fluxes.  The zoom-in box demonstrates the Majorana construction on a tri-coordinate motif and emergent gauge fields $u_{ij}$ corresponding to the flux configuration $\phi_p$.}
    \label{fig:lattice}
\end{figure}

\begin{figure*}[t]
    \centering
    \includegraphics[width=0.9\linewidth]{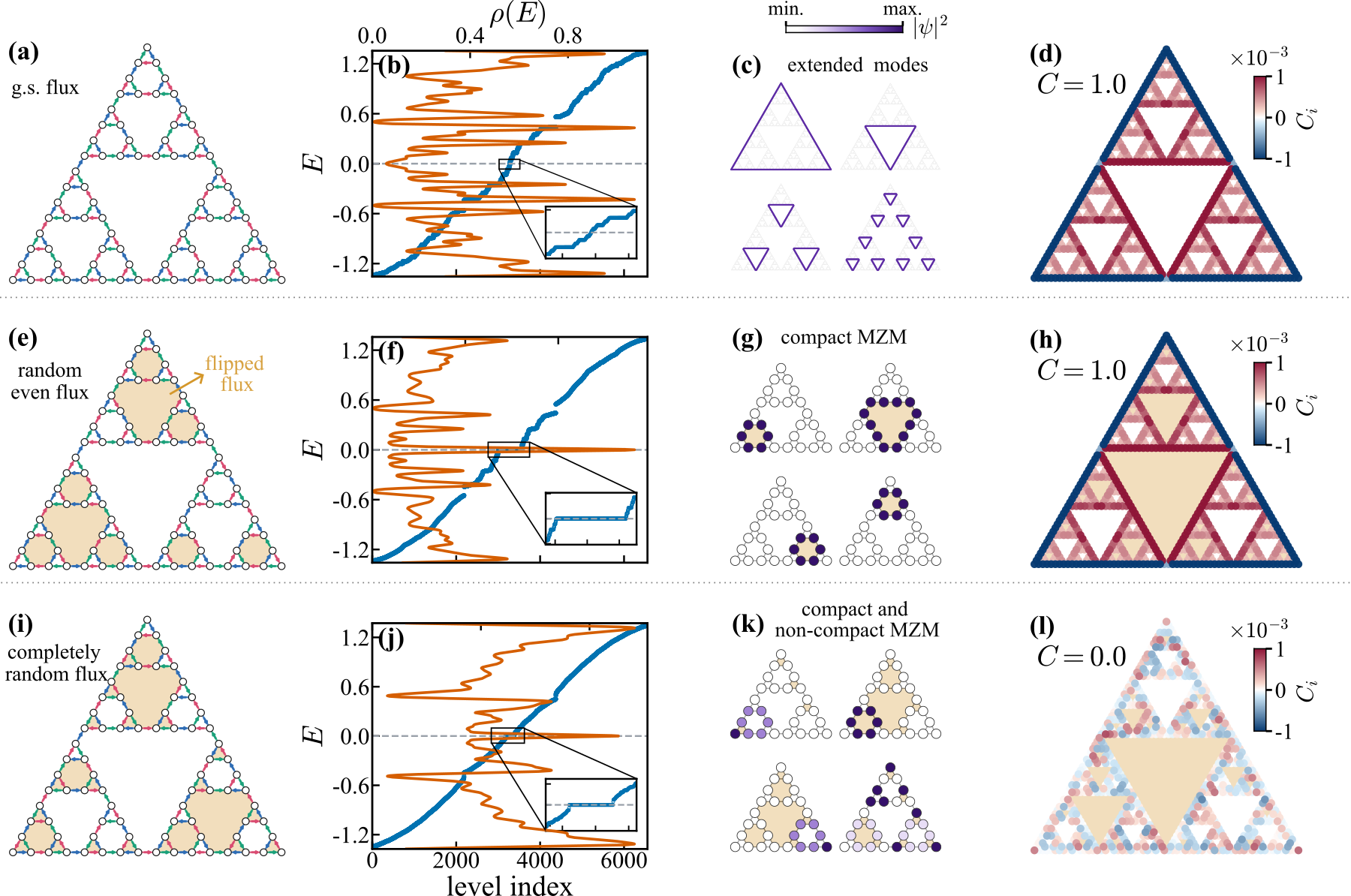}
    \caption{
    \textbf{Summary of fermion sector's response to flux-disorder in the dual-Sierpinski Kitaev model.}
    The three rows show the ground-state flux sector, even-plaquette flux disorder, and full flux disorder respectively.
    (a,e,i) Representative flux sectors; amber plaquettes mark flipped fluxes and arrows indicate the chosen $\mathbb{Z}_2$ gauge orientation.
    (b,f,j) Majorana energy levels and density of states, with insets resolving the near-zero-energy window.
    (c) Schematic hierarchy of boundary-supported extended modes on the outer boundary and successive internal Sierpinski holes.
    (g) Compact MZMs generated by even-plaquette flux flips; colors show normalized local weight $|\psi|^2$.
    (k) Representative MZMs in full flux disorder, including three compact and one non-compact supports.
    (d,h,l) Local Chern marker $C_i$ computed from the real-space projector; the integrated marker remains quantized for the ground-state and even-disordered sectors but is destroyed by full flux disorder.
    }
    \label{fig:everything}
\end{figure*}

In this work, we show that the Kitaev-Sierpinski gasket (KSG) [Fig.~\ref{fig:lattice}] supports a gapless chiral spin liquid with a hierarchy of boundary modes [Fig.~\ref{fig:everything}(a-d)]. Remarkably, an extensive zero-energy degeneracy appears under strong flux disorders [Fig.~\ref{fig:everything}(f,j)], in sharp contrast to the weakly anti-localized Majorana metal found in disordered two-dimensional topological superconductors~\cite{Fulga20}. We further show that, due to the anomalous response to flux disorders, both propagating and localized Majorana modes in KSG are accessible via quench spin correlation on product states with entanglement within individual triangles.

Usually, unresolved gauge fluxes in such short-range entangled states complicate dynamical probes: they generally occupy many random flux sectors \cite{Louk19,GuoYi21}, where flux-bound Majorana zero modes (MZMs) can hybridize, destroying the fractionalized band topology \cite{Huse2012,SelfPRB2019,Zhu2025,Wang2025,Feng2025,Feng2026}. Careful preparation of uniform flux sector and matter vacuum has therefore been an important ingredient in recent proposals and demonstrations of fractionalized chiral dynamics in the quantum simulations of Kitaev spin liquid~\cite{Marin23,Evered2025,ali2025,Park25}.

We find instead that, with uniform triangle chirality in a product state, even-plaquette defects bind compact MZMs (CMZMs) that do not hybridize, allowing chiral propagation to survive extensive flux disorder [see Fig.~\ref{fig:lattice} and Figs.~\ref{fig:everything}(e-h)].  Dynamical dephasing of high-energy modes over these random fluxes, which is intrinsic to quench spin dynamics on short-range entangled initial states, then reveals universal chiral edge dynamics~\cite{companion_work}. The gasket’s fractal connectivity and the initial states’ global spin parity further yield an exact parton representation of corner-spin correlations, enabling directional readout of chiral edge propagation through ordinary two-spin correlations after partial flux preparation (without fermion vacuum optimization). Even when fully random fluxes destroy chiral topology, surviving corner CMZMs produce persistent late-time memory from the corresponding polarized state [Fig.~\ref{fig:everything}(i-l)]]. Fractal connectivity thus shapes the nature of fractionalized excitations and their experimental accessibility, enabling much simpler probes of chiral edge dynamics and quantum memory in synthetic simulators.



\emph{ \color{tum}{Kitaev-Sierpinski Model.--}} 
The KSG is defined on the dual Sierpinski gasket, a self-similar network of triangles and holes repeated across length scales
[Fig.~\ref{fig:lattice}]. Its recursive structure admits an
alternating Kitaev bond coloring.
The three outer corners are two-coordinated; spin components along their missing bond colors provide local couplings to external probes for the dynamical correlations studied below. With a spin-$\tfrac12$ at each vertex, the Hamiltonian is $H=J\sum_{\langle jk\rangle_\gamma}
\sigma_j^\gamma\sigma_k^\gamma$, where $\gamma=x,y,z$ selects the coupled Pauli-spin component. Here we focus on the isotropic Kitaev exchange; anisotropic cases will be discussed later in a companion work \cite{companion_work2}. 
Representing each spin by four Majoranas via
$\sigma_j^\gamma=ib_j^\gamma c_j$ enlarges the Hilbert space \cite{Kitaev2006}.
Physical spin states satisfy the constraint $D_j|\widetilde\Psi\rangle=|\widetilde\Psi\rangle$ at every site, with $D_j=b_j^xb_j^yb_j^zc_j$.
The link operators $\hat u_{\langle jk\rangle_\gamma}=ib_j^\gamma b_k^\gamma$ are conserved and have eigenvalues $u_{\langle jk\rangle_\gamma}=\pm1$.
Fixing these signs reduces the spin problem to free matter Majoranas $c_j$ moving in a static gauge background, $H_u=-iJ\sum_{\langle jk\rangle_\gamma} u_{\langle jk\rangle_\gamma}c_jc_k$.

The clockwise product of bond-spin operators
$\sigma_j^\gamma\sigma_k^\gamma$ around a plaquette $p$ defines the conserved plaquette operator $W_p$, which corresponds to the flux
$\phi_p=\prod_{\langle jk\rangle_\gamma\in\partial p}
(-iu_{\langle jk\rangle_\gamma})$: $\phi_p=\pm1$ on even-sided plaquettes and $\pm i$ on triangles.
Time reversal changes the triangle fluxes, so selecting
one uniform triangle chirality breaks this symmetry
\cite{YaoKivelson2007}.
Note the KSG ground-state flux sector cannot be determined by Lieb's flux theorem due to the lack of lattice symmetry \cite{Lieb1994}.
For our ground-state reference we adopt the rule
$\phi_p^{\rm gs}=-(\pm i)^{n_p}$, supported numerically \cite{supp},
where $n_p$ is the number of plaquette sides. We choose one of the two degenerate ground-state sectors as represented in Fig.~\ref{fig:everything}(a). Holes on all scales preclude the usual bulk-edge separation, producing an infinite hierarchy of chiral channels as fractal level $l\to\infty$. Therefore, chiral modes propagate along both the outer and internal boundaries [Fig.~\ref{fig:everything}(c)], thus its fermion DOS display gapless excitations characteristic of fractal $p$-wave superconductors \cite{supp} [Fig.~\ref{fig:everything}(b)].
The resulting gapless phase nevertheless has a quantized Chern number $|C|=1$ \cite{Bianco11} [Fig.~\ref{fig:everything}(d)]. Remarkably, beyond the ground-state sector, flux disorder offers richer static and dynamic features in KSG, as detailed below.

\emph{ \color{tum}{CMZM and Chiral Modes under Flux Disorder.--}} 
The matter DOS displays a flat-band divergence at $E=0$ [Fig.~\ref{fig:everything}(e,f)] under random static fluxes on even-sided plaquettes with pinned triangle fluxes $\phi_\triangle$ [Fig.~\ref{fig:everything}(e)]. This is attributed to the extensive manifold of even-flux-bound CMZMs due to the destructive interference akin to Aharonov-Bohm-caged compact states \cite{Sutherland1986,Vidal98,Naud01,Hang22,Gloria26,Zhang25,Bauer2026}, whose amplitude vanish everywhere except on sites enclosing plaquettes with flipped fluxes [Fig.~\ref{fig:everything}(g)], as detailed in Supplemental Material \cite{supp}, implying that the defect-bound MZMs do not hybridize into a trivial Majorana metal. This gives rise to the extensive degeneracy at zero energy, in sharp contrast to the honeycomb model \cite{Huse2012,Zhu2025}. Further information on the localization of disordered KSG are detailed in \cite{supp}. Due to the lack of MZM hybridization, the chiral boundary modes in KSG remains robust under even-plaquette flux disorder, as confirmed by the $C=1$ Chern number [Fig.~\ref{fig:everything}(h)] \footnote{Since the compact zero-mode subspace is degenerate, including only an arbitrary subset of it would make the energy projector $P$ of Chern marker basis-dependent.  We therefore use one of the two canonical choices: $P=P_-$ or $P=P_-+P_0$, where $P_-$ projects onto all negative-energy states and $P_0$ projects onto the entire exact zero-mode subspace. Both give consistent results. See also \cite{supp}. }. We emphasize that $\phi_\triangle$ must be uniform, since random fluxes on both triangles and even-plaquette faces [Fig. \ref{fig:everything}(i)] can generate extensive zero modes and the resulting MZM hybridization trivializes the topology [Figs.~\ref{fig:everything}(i-l)].

\begin{figure*}[t]
    \centering
    \includegraphics[width=0.9\linewidth]{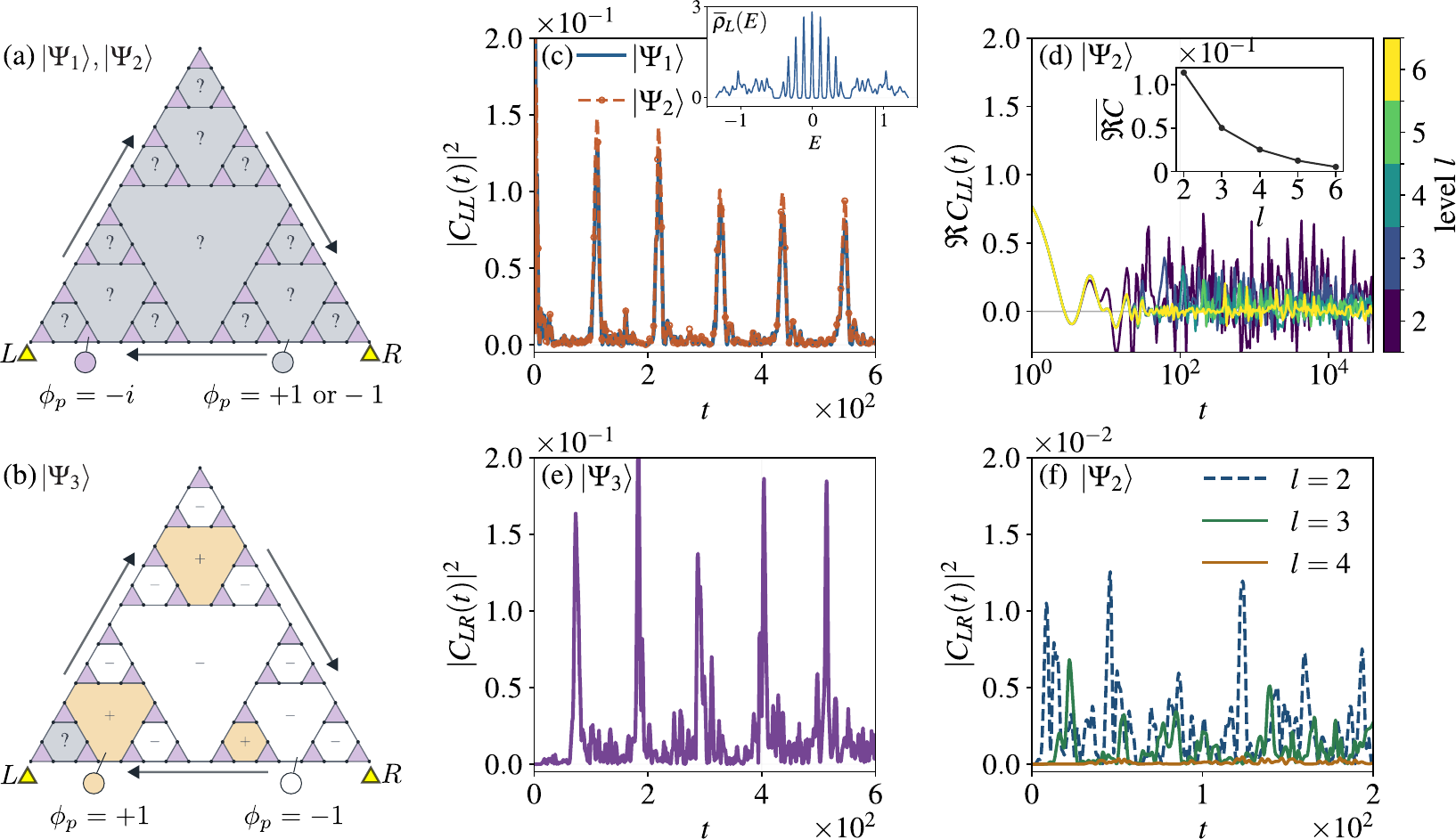}
    \caption{\textbf{Flux sectors of the initial states and results of spin quench dynamics.}
    (a) Flux of the triangular product states $\ket{\Psi_{1(2)}}= \bigotimes_\triangle\lvert \triangle_{1(2)}\rangle$: triangular fluxes are pinned to $-i$, while even-sided plaquettes remain unresolved.
    (b) Flux sector of $\ket{\Psi_3}$, where triangular fluxes are pinned to $-i$ and all but one selected even-sided plaquette are projected to random flux eigen values. Arrows are indicators of chiral edge modes. 
    (c) Intermediate-time dynamics of the onsite corner correlator $|C_{LL}(t)|^2$ for $\ket{\Psi_1}$ and $\ket{\Psi_2}$ at $l=4$. Inset shows the disorder-averaged onsite local DOS. 
    (d) Late-time real part $\mathfrak{R}\,C_{LL}(t)$ for $\ket{\Psi_2}$ at multiple fractal levels $l$. Inset: corresponding late-time averages versus $l$.
    (e) Cross-corner correlator $|C_{LR}(t)|^2$ for $\ket{\Psi_3}$ at $l=4$, with a single unfixed even flux (gray) and quenched randomness in remaining even fluxes, as shown in (b).
    (f) Cross-corner correlator $|C_{LR}(t)|^2$ for $\ket{\Psi_2}$, as shown in (a), at several fractal levels $l$, illustrating strong dephasing with increasing system size; the $l=2$ curve is rescaled by $1/50$ for visibility. All curves are averaged by $1000$ samples.
    }
    \label{fig:correlator_dynamics}
\end{figure*}

We discuss two exemplary initial product states giving even-plaquette flux disorder in the fractionalized basis: $|\Psi_1\rangle$ and $|\Psi_2\rangle$, as products over disjoint elementary triangles $|\Psi_{1(2)}\rangle = \prod_\triangle |\triangle_{1(2)}\rangle$, with $|\triangle_{1(2)}\rangle$ having a fixed triangular flux $\phi_\triangle$ and a definite three-spin parity $\prod_{j\in \triangle}\sigma_j^z$, leaving all even-plaquette fluxes unresolved  [Fig.~\ref{fig:correlator_dynamics}(a)]. For $|\Psi_1\rangle$, each triangle is prepared in a ground state of its three-spin Hamiltonian with flux $\phi_\triangle=-i$ and even spin parity. For $|\Psi_2\rangle$, the same triangle flux is selected by a local measurement starting from a $z$-polarized state. See End Matter for their preparation.
In the fractionalized basis, these product states are coherent superpositions of even-plaquette flux disorders, $|\Psi\rangle=\sum_{\boldsymbol w}Q_{\boldsymbol w}|\Psi\rangle$, where $Q_{\boldsymbol w}$ projects onto an even-plaquette
flux configuration $\boldsymbol w$. Every nonzero component retains the same triangle flux $\phi_\triangle=-i$. The even-flux defects bind CMZMs rather than hybridizing into extended zero-energy states \cite{Huse2012}, preserving
the Chern character of the matter Hamiltonians [Fig.~\ref{fig:everything}(e-h)]. The preparation therefore accesses an ensemble of chiral Majorana copies without selecting a flux-free pattern or preparing their fermion vacuum \cite{supp}.

\emph{\color{tum}{Non-Equilibrium Spin Dynamics.--}}
Now we show that the fractal geometry and the symmetry of the aforementioned initial product states enable an exact and closed parton representation of chiral spin dynamics. 
The Kitaev spin fractionalization enables the evaluation of $\langle \Psi |\hat{O}(t)| \Psi \rangle$ in the matter sector as long as $\hat{O}(t)$ preserves all bond-fermion numbers \cite{Baskaran2007}, a property that enabled the computation of several spectroscopic responses \cite{Knolle2014, Knolle2015, KnolleRaman2014, Halasz2016, QiangPRL2024, Feldmeier2020, Shang-ShungZhang2025} and explains the ultrashort bulk spin correlations \cite{Baskaran2007}. On the other hand, correlations between spin operators at corners can be long-ranged from flux conservation \cite{Feldmeier2020,Shang-ShungZhang2025}. This result applies to the corner-corner dynamical correlations on the KSG 
\begin{equation}
C_{LV}(t)
= \langle \Psi|\,\sigma_V^{\gamma_V}(t)\,\sigma_L^{\gamma_L}(0)\,|\Psi\rangle,
\label{eq:CRL_main}
\end{equation}
in which $L$ is the left corner, $V$ any two-coordinated corner, and $\gamma_{L,V}$ is the axis associated with its absent third bond.

The corresponding $b$ Majorana on corner site $L$ or $V$ belongs to no physical link. The spin probe consequently
leaves all plaquette fluxes unchanged. For $\gamma_L=y$ and
$\gamma_V=x$, substituting $\sigma_j^\gamma=i b_j^\gamma c_j$
into Eq.~\eqref{eq:CRL_main} thus gives the gauge-invariant
four-Majorana correlator $\langle b_V^x b_L^y c_V(t)c_L\rangle$
within each static gauge sector $u$. Crucially, the fractal geometry and the initial states' symmetry
$P_z|\Psi_{1,2}\rangle=|\Psi_{1,2}\rangle$, with
$P_z=\prod_j\sigma_j^z$, allow the dangling pair $b_V^x b_L^y$ to be eliminated from this matrix element in favor of the string factor $\chi_{LV}^{(u)}\propto-i\prod_{\ell\in\{x,y\}}u_\ell$, with the product runs over $x/y$ links within the fractal gasket (End Matter). 
This prevents cancellation between the empty and occupied states of the boundary fermion formed by the dangling pair $b_V^x b_L^y$, enabling the off-site spin correlation as a probe of chiral direction, instead of the more demanding out-of-time-ordered correlations for regular lattices, see our companion work~\cite{companion_work}.
Combining this reduction with Majorana propagator gives our central result for the exact parton representation of the non-equilibrium spin dynamics \cite{supp}:
\begin{equation}
 C_{LV}(t)=\sum_{u,k}p(u)\chi_{LV}^{(u)}
             [e^{A(u)t}]_{Vk}G_{kL}^{(u)}.
 \label{eq:C_fixed_sector}
\end{equation}
Here $e^{A(u)t}$ is the matter-Majorana propagator conditioend on the static gauge field $u$, with $H_u=(i/4)\sum_{mn}A_{mn}(u)c_mc_n$; the probability $p(u)$ of $u$ field and the zero-time fermion correlations $G_{kL}^{(u)}$ given by
\begin{equation}
p(u)=\langle\widetilde\Psi|\Pi_u|\widetilde\Psi\rangle,\;\;G_{kL}^{(u)}=
 p^{-1}(u)\langle\widetilde\Psi|\Pi_u c_kc_L|\widetilde\Psi\rangle.
 \label{eq:chi_reduction}
\end{equation}
where $|\widetilde\Psi\rangle$ is the normalized physical Majorana embedding of the full spin state satisfying $D_j|\widetilde\Psi\rangle=|\widetilde\Psi\rangle$, and $\Pi_u$ the projector onto physical-link eigenvalues $u$. 
End Matter derives this exact reduction, and benchmarks against exact diagonalization (ED).
For the on-site correlation ($V=L$), $(b_L^{\gamma_L})^2=1$
gives $\chi_{LL}^{(u)}=1$. Since even-flux disorder preserves the topology, $C_{LL}(t)$ reveals recurrences of gapless edge modes winding around the gasket, while disorder averaging dephases nonuniversal noises, which will be elaborated in the next section.

\emph{\color{tum}{Chiral Majorana Modes at Intermediate Timescales.--}} The corner correlator introduced above directly tracks the propagation of matter Majorana fermions in the initial state, where the boundary-supported component is biased toward a definite chiral direction. Since the prepared product state is non-equilibrium, one might expect a noisy background masking the chiral signal. However, $C_{LV}(t)$ involves an average of the Majorana correlator over the unresolved even-sided flux disorders, which efficiently dephases non-universal contributions while leaving the chiral recurrence sequence sharply resolved within early and intermediate timescales \cite{companion_work}. We now elaborate on the gauge-invariant dynamical correlator starting from the aforementioned initial product states $|\Psi_{1(2)}\rangle$.

Figure~\ref{fig:correlator_dynamics} summarizes the nonequilibrium corner-spin response on a level-$4$ KSG.  The dynamical correlations are computed from Eq.~\eqref{eq:C_fixed_sector} by averaging over
configurations of the unfixed even-sided plaquettes, while all elementary-triangle chiralities are pinned in the
initial state. Fig.~\ref{fig:correlator_dynamics}(c) displays the onsite corner correlator $|C_{LL}(t)|^2$ for $|\Psi_1\rangle$ and $|\Psi_2\rangle$. 
Both show a pronounced sequence of temporally separated pulses due to the winding chiral mode. Note that Eq.~\eqref{eq:C_fixed_sector} averages amplitudes rather than probability, odd winding thus acquires a relative $\mathbb Z_2$ sign determined by the parity of the enclosed flux flips. Averaging over both parities suppresses odd-winding returns, whereas even windings square this sign to unity, therefore the dominant recurrence spacing is $2T_\partial$, where $T_\partial$ is the time for one complete outer-boundary winding. A weak disorder dependence of the chiral edge dispersion and the delocalization along the boundary produce a slowly decaying envelope, as is also visible in the peak broadening of the disorder-averaged local density of states $\bar{\rho}_L$ in the inset of Fig.~\ref{fig:correlator_dynamics}(c).
In the long-time limit, the chiral mode delocalizes completely along the boundary, reducing the correlation intensity and converging to zero as fractal level $l$ gets higher, as shown in Fig.~\ref{fig:correlator_dynamics}(d).

Note that even though the recurrent pulses in $C_{LL}(t)$ establish the presence of a propagating edge mode, it alone does not determine the chiral direction; we therefore also consider the arrival signal at a distinct corner, e.g.
$C_{LR}(t)$. Here it is imperative to point out a caveat regarding off-diagonal dephasing in $C_{LR}(t)$. The propagator in Eq.~\eqref{eq:C_fixed_sector} has the eigenmode expansion $[e^{A(u)t}]_{Rk}=\sum_n
e^{-i\varepsilon_n^{(u)}t}\varphi_{n,R}^{(u)}\varphi_{n,k}^{(u)*}$,
where $\varepsilon_n^{(u)}$ and $\varphi_{n,j}^{(u)}$ are the energy and site-$j$ amplitude of a normalized eigenmode of $iA(u)$. The $L$ to $R$ propagation contains the complex overlap $\varphi_{n,R}^{(u)}\varphi_{n,L}^{(u)*}$, which can differ in phase between different flux sectors, making disorder averaging also dephase the chiral signal in addition to the nonuniversal noise background. 
This can be seen from Fig.~\ref{fig:correlator_dynamics}(f), where correlation between different corners $C_{LR}(t)$ for $|\Psi_{2}\rangle$ presents strong dephasing as the fractal level increases, rapidly converging to zero.
This unwanted off-diagonal dephasing, however, can be mitigated by preparing a $P_z$-preserving $|\Psi_3\rangle$ with partially resolved random even-plaquette fluxes, see Fig.~\ref{fig:correlator_dynamics}(b), which can be easily made by an ancilla protocol based on $\ket{\Psi_{1(2)}}$ (End Matter). The narrower random flux ensemble therein provides strong enough even-plaquette randomness for dephasing the nonuniversal high-energy noise,  while weak enough for retaining the directional chiral signals in $C_{LR}(t)$, as shown in Fig.~\ref{fig:correlator_dynamics}(e). 
Here we point out that perturbations and Trotter errors in digital simulation can spoil the interference that confines CMZMs, compromising chiral dynamics. This, however, can be resolved by an engineered Trotterization that preserves CMZMs and chiral topology, which will be detailed in a companion work \cite{companion_work}.

\emph{\color{tum}{Late-time Signature of Localized CMZM.--}}
Now we discuss the late-time dynamics of $C_{LL}(t)$ from the fully polarized initial state $\lvert\Uparrow\rangle=\bigotimes_j\lvert0_z\rangle_j$, which fractionalizes into the fully-disordered flux ensemble [Fig.~\ref{fig:everything}(i)], allowing MZMs' hybridization; nevertheless, some CMZM still survives, as shown in Fig.~\ref{fig:everything}(k) where compact and non-compact zero modes coexist, resulting in a weaker DOS divergence at $E=0$ shown in Fig.~\ref{fig:everything}(j).
Enabled by the fractal structure, the surviving CMZM here spans a very small spatial extent near corners compared to similar compact zero modes in other regular lattice geometries \cite{Bauer2026}, indicating a sharp late-time dynamical signature from the corner CMZM [Fig.~\ref{fig:everything}(k)]. 

\begin{figure}[t]
    \centering
    \includegraphics[width=0.9\linewidth]{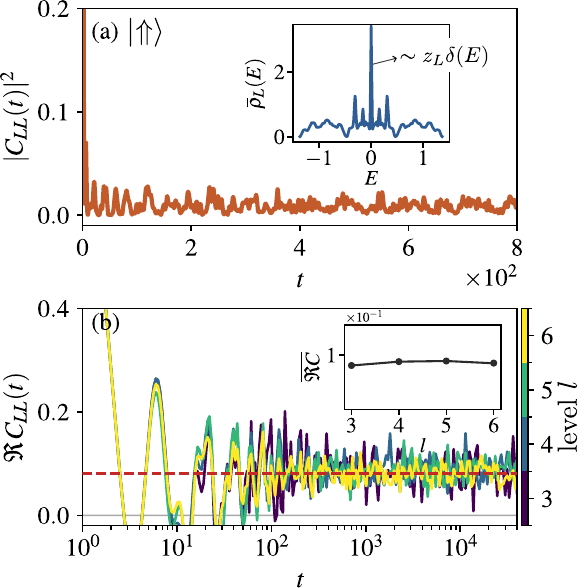}
    \caption{\textbf{Polarized-state dynamics of KSG.}
    (a) Intermediate-time onsite correlator $ |C_{LL}(t)|^2 $ computed from Eq.~\eqref{eq:C_fixed_sector} for the polarized initial state in a $l=4$ gasket.
    Inset: Averaged corner local density of states $\rho_L(E)$ over fully random flux disorder.
    (b) Long-time real part $\mathfrak{R}\,C_{LL}(t)$ for multiple fractal levels $l$; the horizontal red line marks the late-time average.
    Inset: late-time value versus fractal level $l$, showing the size independence of the long-time remnant.
    }
    \label{fig:correlator_dynamics_polarized}
\end{figure}

For a fully polarized state, the Fourier-transformed averaged local DOS $\bar{\rho}_L$ is directly the averaged return amplitude $[e^{A(u)t}]_{LL}$; correspondingly, the disorder-averaged correlator inherits the low-frequency structure of the local spectral measure at the corner.  
We write the $C_{LL}(t)$ in spectral form: $C_{LL}(t)
\sim\int \mathrm{d}E\,\overline{\rho}_L(E)\,e^{-iE t}$ with $\overline{\rho}_L(E)=\overline{\rho}_L^{\rm reg}(E)+z_L\,\delta(E)$. 
Here the overline denotes the average over the sampled static flux sectors, so $\overline{\rho}_L(E)$ is the corresponding flux-ensemble-averaged local spectral weight at site $L$, and $z_L$ denotes the strength of the zero-energy singularity due to the divergent local DOS of the CMZM at the corner $L$ [inset of Fig.~\ref{fig:correlator_dynamics_polarized}(a)]. 
The non-topological nature of $\lvert\Uparrow\rangle$ [Fig.~\ref{fig:everything}(l)] and a relatively smooth $\overline{\rho}_L(E\neq 0)$ produces a $C_{LL}(t)$ that decays rapidly as $1/\rm bandwidth$ without recurrent pulses [Fig.~\ref{fig:correlator_dynamics_polarized}(a)], in sharp contrast to that of the chiral initial state in Fig.~\ref{fig:correlator_dynamics}(c). By contrast, a divergent $\overline{\rho}_L(E=0)$ leads to a stable late-time plateau shown in Fig.~\ref{fig:correlator_dynamics_polarized}(b), with $C_{LL}(t\to \infty) \simeq z_L$ finite.

\emph{\color{tum}{Conclusions and Outlook.--}}
The KSG supports a gapless fractal chiral spin liquid whose topological matter sector survives extensive even-plaquette flux disorder with fixed triangle chirality. The flux-bound zero modes remain strictly compact rather than hybridizing into a topologically trivial Majorana metal \cite{Huse2012}. The fractal geometry and $P_z$-preserving triangle-product initial states together enable an exact parton representation of non-equilibrium corner-spin correlation, providing an exact and sharp access to its rich topological physics dynamically. Physical spin correlations reveal chiral boundary modes' recurrences, while partial even-flux measurements recover directional corner-to-corner pulses, and fully polarized states exhibit a late-time remnant from corner-localized CMZMs. Fractal geometry thus enables complementary probes of mobile and localized compact Majorana excitations without preparing the many-body ground state nor an energetically optimal flux sector. 
Experiments on chiral transport in fractal photonic lattices
\cite{Biesenthal2022,LiFractalFloquet2023}, interacting Rydberg atoms on Sierpinski arrays \cite{verstraten2025}, and topological states and Kitaev dynamics in quantum simulators \cite{Semeghini21,Iqbal2024,Evered2025} provide key experimental ingredients for pursuing this program. 
Gapless chiral spin liquids have also been recently constructed in hyperbolic lattices \cite{Lenggenhager2025,Dusel2025}, where the quench dynamics for quantum simulators in relation to their fractionalized excitaions can be studied by the same formalism.

\medskip
\textbf{Methods}

\begin{figure}[t]
    \centering
    \includegraphics[width=0.85\linewidth]{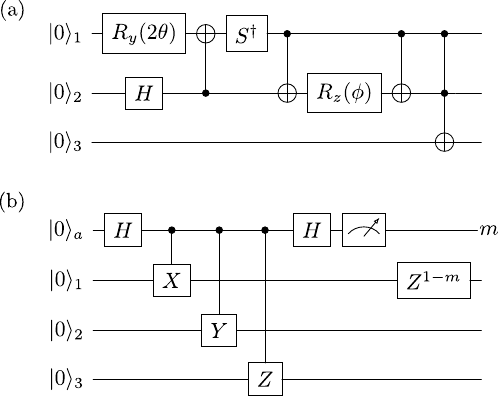}
    \caption{\textbf{Local quantum circuits for initial state preparation.} (a) Unitary circuit that initializes the $|\triangle_{1}\rangle$ state. Rotation angles as set by $\theta = \cos^{-1}\left(\frac{\sqrt{6(3-\sqrt{3})}}{6} \right)$ and $\phi = 3\pi/4$.  (b) Circuit that initializes $|\triangle_{2}\rangle$ by entangling an ancilla qubit to the fractal sites followed by a single measurement.}
    \label{fig:init_circuit}
\end{figure}

{\it \color{tum} Appendix A: Preparation for initial states.---}
We present two ways to prepare chiral-biased initial states with a triangle-product structure: one via purely unitary gates, and the other using measurement-based non-unitary preparation. Let a single-triangle energy be
\begin{equation}
H_{3}=\sigma_1^z \sigma_2^z + \sigma_2^x \sigma_3^x + \sigma_3^y \sigma_1^y.
\label{eq:H3_spin_EM}
\end{equation}
In terms of Majorana fermions, the ground state obtained for fixed $\phi_\triangle = -i$ reads
\begin{equation}
\begin{split}
|\triangle_{1}\rangle
&=\frac{\sqrt{3(3-\sqrt{3})}}{6}\,|000\rangle
+e^{i\frac{3\pi}{2}}\frac{\sqrt{3(3-\sqrt{3})}}{6}\,|110\rangle \nonumber\\
&\hspace{0.2em}
+e^{i\frac{3\pi}{4}}\frac{\sqrt{3(3+\sqrt{3})}}{6}\,|011\rangle +e^{i\frac{\pi}{4}}\frac{\sqrt{3(3+\sqrt{3})}}{6}\,|101\rangle,
\end{split}
\label{eq:psi3_spin_EM}
\end{equation}
which satisfies $P_z|\triangle_{1}\rangle=|\triangle_{1}\rangle$. This state can be initialized using the unitary circuit indicated in Fig. \ref{fig:init_circuit}(a). Applying this quantum circuit in parallel for all triangles on the fractal initializes the state $|\Psi_1\rangle$ with $P_z = +1$ discussed in the main text.

Alternatively, we can fix the flux from any initial state following the nonunitary protocol detailed in Ref. \cite{ali2025}. As indicated in Fig. \ref{fig:init_circuit}(b), we need to introduce an ancilla qubit $a$, apply the Hadamard gate, entangle the qubit with the fractal through control gates, and reapply the Hadamard gate. For a single triangle, this circuit performs the transformation 
\begin{equation}
    |0_a\rangle\lvert \Uparrow\rangle \rightarrow \frac12 \left[|0_a\rangle(1-i W_\triangle) + |1_a\rangle (1+iW_\triangle) \right]\lvert \Uparrow\rangle.
\end{equation}
where $\lvert \Uparrow\rangle = \bigotimes_j\lvert0_z\rangle_j$ is the fully polarized product state along $\hat{z}$. 
Measuring the ancilla qubit fixes the triangular fluxes, thus providing the protocol to initialize $|\Psi_2\rangle$: 
starting from $|000\rangle$, either retain $m=1$ or apply
$Z_1^{1-m}$ to obtain, for both outcomes,
\begin{equation}
 |\triangle_{2}\rangle=\frac{(1+iW_\triangle)|000\rangle}{\sqrt2}
 =\frac{|000\rangle-i|110\rangle}{\sqrt2}.
 \label{eq:psi2_spin_EM}
\end{equation}
with a fixed triangular flux. 
Thus $|\Psi_2\rangle=\bigotimes_\triangle|\triangle_{2}\rangle$
has the same triangle flux. Both local states contain only zero or two down spins, so
$P_z=\prod_\triangle(\sigma_1^z\sigma_2^z\sigma_3^z)$ gives
$P_z|\Psi_{2}\rangle=|\Psi_{2}\rangle$.

The same ancilla protocol measures selected even-plaquette
stabilizers on either $\ket{\Psi_1}$ or $\ket{\Psi_2}$ to prepare $|\Psi_3\rangle$, while preserving the commuting
triangle fluxes.
For a nonzero-probability record $\boldsymbol w$,
\begin{align}
 Q_{\boldsymbol w}&=\prod_{p\ {\rm measured}}\frac{1+w_pW_p}{2},\\
 |\Psi_3\rangle&=\frac{Q_{\boldsymbol w}|\Psi_a\rangle}
 {\|Q_{\boldsymbol w}|\Psi_a\rangle\|},\;\;a=1,2.
\end{align}
Here $w_p=\pm1$ are the recorded even-flux eigenvalues.
Conjugation by $P_z$ reverses each $\sigma^{x,y}$, so the two signs cancel on every same-color bond. Every ordered plaquette product therefore commutes with $P_z$, as does $Q_{\boldsymbol w}$. It follows that $P_z|\Psi_3\rangle=+|\Psi_3\rangle$ remains unchanged from its parental state $\ket{\Psi_1}$ or $\ket{\Psi_2}$.

\medskip
\begin{figure*}[t]
 \centering
 \includegraphics[width=0.95\textwidth]{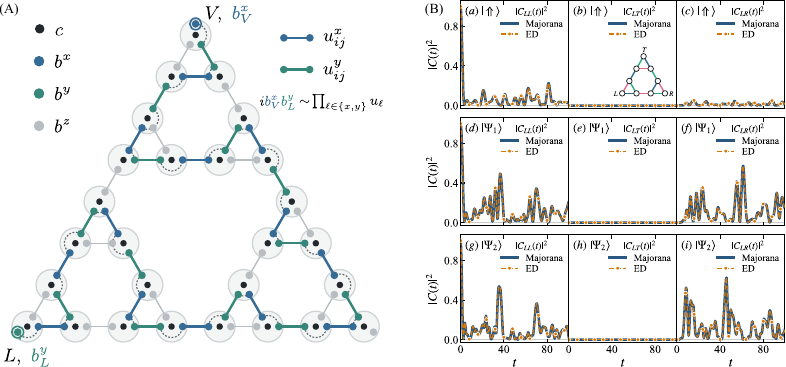}
 \caption{\textbf{Left: Corner partons on the level-three KSG.}
 Each disk contains $c,b^x,b^y,b^z$; solid links pair gauge
 Majoranas and dashed arcs are eye-guiding lines completing the all-site covering $x/y$ string.
 Fixed $P_z$ and resolved links determine the ringed $L,V$
 pair's parity. The third dangling $b^z$ does not enter this identity hence denoted by gray bullets.
 Right: 
 ED vs Majorana averaging benchmark for level-2 KSG, shown in the inset of (b). Rows correspond to initial states $\lvert0\rangle$, $|\Psi_1\rangle$, and $|\Psi_2\rangle$; columns show $|C_{LL}(t)|^2$, $|C_{LT}(t)|^2$, and $|C_{LR}(t)|^2$. ED and Majorana averaged results overlap panel-by-panel. In (b,e,h) of (B), $C_{LT}(t)=0$ because $\sigma_T^z(t)\sigma_L^y$ is odd under $P_z$, while all three initial states are even under $P_z$.
 }
 \label{fig:parity_path}
\end{figure*}

{\it \color{tum} Appendix B: Derivation of Eq.~\eqref{eq:C_fixed_sector}.-- }
Here we present the exact parton representation for the corner-spin correlations for the three initial states used in the main text.  Use the normalized physical embedding of the full spin state, $|\widetilde\Psi\rangle$, with $D_j|\widetilde\Psi\rangle=|\widetilde\Psi\rangle$,
$D_j=b_j^xb_j^yb_j^zc_j$.
Physical links can be resolved with the projector:
\begin{equation}
    \Pi_u=\prod_\ell(1+u_\ell\hat u_\ell)/2,\;\;\;p(u)=\langle\widetilde\Psi|\Pi_u|\widetilde\Psi\rangle.
\end{equation}
The corner spin is denoted by $\sigma_j^{\gamma_j}=i b_j^{\gamma_j}c_j$, which preserves every
physical link, and its dangling $b_j^{\gamma_j}$, as shown in Fig.~\ref{fig:parity_path}(A) for sites $L$($V$) and color $\gamma_L=y$($\gamma_V=x$), commutes with $H$. Consequently, for a given gauge sector,
\begin{equation}
 \Pi_u \sigma_V^{\gamma_V}(t)\sigma_L^{\gamma_L} \Pi_u
 = b_V^{\gamma_V}b_L^{\gamma_L}c_V(t;u)c_L.
 \label{eq:C_fixed_sector1_app}
\end{equation}
Inserting $\sum_u\Pi_u=1$ and $c_V(t;u)=\sum_k[e^{A(u)t}]_{Vk}c_k$ gives the physical
spin correlator defined in the main text:
\begin{equation}
 C_{LV}(t)=\sum_{u,k}p(u)[e^{A(u)t}]_{Vk}C_{kL}^{(u)}(0).
 \label{eq:AppA_propagation_input}
\end{equation}
Here the zero-time coefficient appearing in Eq.~\eqref{eq:chi_reduction} is
\begin{equation}
 C_{kL}^{(u)}(0)=p^{-1}(u)\langle\widetilde\Psi|\Pi_u b_V^{\gamma_V}b_L^{\gamma_L}
  c_kc_L|\widetilde\Psi\rangle.
 \label{eq:C_fixed_sector2_app}
\end{equation}
If the dangling parity $ib_V^{\gamma_V}b_L^{\gamma_L}=\pm1$ is unresolved, the two parity sectors enter with opposite signs and cancel completely for any initial state \cite{Feldmeier2020}.

This formal parity cancellation, however, can be gauged away by
combining the global symmetry of the initial states with the gasket's fractal geometry: both prepared states have definite global spin parity, $P_z|\widetilde\Psi\rangle=q_z|\widetilde\Psi\rangle$, with
$q_z=+1$ for $|\widetilde\Psi\rangle = |\Psi_{1,2,3}\rangle$ (see Appendix A); and the dual Sierpinski gasket has a single $x/y$ path through all $N=3^l$ sites, as shown by the alternating blue and green bonds in Fig.~\ref{fig:parity_path}(A), with exactly these two dangling endpoints.

To see this explicitly, in the following we use $\gamma_L = y$ and $\gamma_V = x$ as an example. The local identity $\sigma_j^zD_j=-i b_j^xb_j^y$, physicality condition, and pairing along the $x/y$ path in Fig.~\ref{fig:parity_path}(A) give
\begin{equation}
 \begin{aligned}
 |\widetilde\Psi\rangle
 =\Bigl[\prod_j(-i b_j^xb_j^y)\Bigr]|\widetilde\Psi\rangle
 = \eta i\,b_V^x b_L^y\prod_{\ell\in\{x,y\}}\hat u_\ell |\widetilde\Psi\rangle.
 \end{aligned}
 \label{eq:AppA_Pz_projected}
\end{equation}
The sign $\eta=\pm1$ is fixed by site ordering and bond
orientations. Apply $\Pi_u$ to
Eq.~\eqref{eq:AppA_Pz_projected} and use the commutation relation
$\Pi_u\hat u_\ell=u_\ell\Pi_u$:
\[
 \Pi_u|\widetilde\Psi\rangle
 =\eta i\Bigl(\prod_{\ell\in\{x,y\}}u_\ell\Bigr)
          b_V^x b_L^y\Pi_u|\widetilde\Psi\rangle.
\]
Since $u_\ell^2=\eta^2=1$, division by the scalar prefactor yields
\begin{align}
 b_V^x b_L^y\Pi_u|\widetilde\Psi\rangle
 =q_z\chi_{LV}^{(u)}\Pi_u|\widetilde\Psi\rangle,\;\;
 \chi_{LV}^{(u)}=-i\eta\prod_{\ell\in\{x,y\}}u_\ell.
 \label{eq:AppA_bb_projected_final}
\end{align}
Finally, plugging Eq.~\eqref{eq:AppA_bb_projected_final} into Eq.~\eqref{eq:C_fixed_sector2_app} gives the desired result in Eq.~\eqref{eq:chi_reduction}, thus Eq.~\eqref{eq:C_fixed_sector} 
for nonequivalent corners. This is a feature directly tied to Sierpinski gasket's geometry and the symmetry therein, and is not available in the dynamical protocol for regular spin liquid systems where out-of-time-ordered correlator (OTOC) is required for a non-trivial off-site correlator \cite{companion_work}. For the correlation on the same corner $V=L$, on the other hand, $(b_L^{\gamma_L})^2=1$ instead gives $C_{kL}^{(u)}(0)=G_{kL}^{(u)}$ directly without the involvement of dangling $b$ fermions.

Every static link configuration $u$ is compatible with a physical spin state: boundary zero modes allow the fermion-parity constraint to be satisfied \cite{Pedrocchi11,Zschocke2015}. On the gasket, the unused corner $b^z$ leaves this freedom even after the selected pair parity is fixed. For each configuration, we evolve the Majorana operators with $e^{A(u)t}$, combine this evolution with
the initial correlations $G_{kL}^{(u)}$ and parity factor
$\chi_{LV}^{(u)}$, and average with the prepared state's
probabilities $p(u)$, as in Eq.~\eqref{eq:C_fixed_sector}. This sum gives the exact spin correlation, see Fig.~\ref{fig:parity_path}(B).

The triangle-product state's definite $P_z$ and the gasket's
single $x/y$ path together fix the selected dangling-pair parity, removing its independent cancellation. Global parity alone does not generally fix such a pair on e.g. the star-lattice edge. This does not, however, eliminate dephasing between physical flux sectors: for $V\neq L$, the complex overlaps $\varphi_{n,L}^{(u)*}\varphi_{n,V}^{(u)}$
contribute sector-dependent phases to the spin response. Measuring selected even-plaquette fluxes reduces the unresolved ensemble and controls this dephasing, revealing the directional peaks in
Fig.~\ref{fig:correlator_dynamics}(b,e). The measured fluxes may be random; neither flux correction nor a matter vacuum is needed.

\medskip
{\it \color{tum}{Data availability.-- }}
Data are available upon reasonable request on Zenodo~\cite{zenodo}.

\medskip
{\it \color{tum}{Author Contribution.-- }}
S.F., J.K., and W.N. jointly conceived the project. S.F. and W.N. conducted the analytical study. S.F. performed the calculations. All authors contributed to writing the manuscript.

\medskip
\acknowledgments{
{\it \color{tum}{Acknowledgments.-- }}
We are grateful to  P. d'Ornellas, F. Dusel, K. Kobayashi, M. Knap, F. Pollmann, and R. Pereira for discussions.   
S.F. and J.K. acknowledge support from the Deutsche Forschungsgemeinschaft (DFG, German Research Foundation) under Germany’s Excellence Strategy--EXC--2111--390814868, TRR 360 – 492547816 and DFG grants No. KN1254/1-2, KN1254/2-1, the Imperial-TUM flagship partnership, the European Research Council (ERC) under the European Union’s Horizon 2020 research and innovation programme (grant agreement No. 851161), the European Union (grant agreement No. 101169765), as well as the Munich Quantum Valley, which is supported by the Bavarian state government with funds from the Hightech Agenda Bayern Plus. S.F. was also supported in part by grant NSF PHY-2309135 to the Kavli Institute for Theoretical Physics (KITP) during the KITP Program: Engineering Fractionalization in Quantum Many-Body Systems -- From New Quantum Materials to Quantum Simulators. W. N. acknowledges support from Conselho Nacional de Desenvolvimento Científico e Tecnológico (CNPq) for support through productivity grant 305739/2024-7 and the support of the INCT project Advanced Quantum Materials, involving the Brazilian agencies CNPq (Proc. 408766/2024-7), FAPESP (Proc. 2025/27091-3), and CAPES.
}

\let\oldaddcontentsline\addcontentsline
\renewcommand{\addcontentsline}[3]{}

\bibliography{Fractals}

\end{document}


\title{Supplemental Material for ``\textbf{\ourtitle}"}

\author{Shi Feng}
\email{shi.feng@tum.de}
\affiliation{Technical University of Munich, TUM School of Natural Sciences, Physics Department, 85748 Garching, Germany}
\affiliation{Munich Center for Quantum Science and Technology (MCQST), Schellingstr. 4, 80799 M{\"u}nchen, Germany}

\author{Johannes Knolle}
\affiliation{Technical University of Munich, TUM School of Natural Sciences, Physics Department, 85748 Garching, Germany}
\affiliation{Munich Center for Quantum Science and Technology (MCQST), Schellingstr. 4, 80799 M{\"u}nchen, Germany}
\affiliation{Blackett Laboratory, Imperial College London, London SW7 2AZ, United Kingdom}
\author{Willian M. H. Natori}
\affiliation{Gleb Wataghin Institute of Physics, University of Campinas, Campinas,
São Paulo 13083-950, Brazil}

\maketitle
\setcounter{section}{0}
\setcounter{figure}{0}
\setcounter{equation}{0}
\renewcommand{\thefigure}{S\arabic{figure}}
\renewcommand{\theequation}{S\arabic{equation}}
\renewcommand{\thesection}{S\arabic{section}}
\onecolumngrid

This Supplemental Material develops the spectral and dynamical analysis underlying the main text. Section~\ref{sec:SM_ground} tests the reference ground-state flux sector and derives a recursive Green's function for the Kitaev--Sierpinski gasket (KSG). Sections~\ref{sec:SM_ipr} and \ref{sec:SM_wavepacket} use eigenstate participation ratios and real-space propagation to distinguish localized weight from modes circulating along the outer and internal boundaries. Section~\ref{sec:SM_spin} then derives the exact representation of corner-spin correlations in terms of Majorana propagation. Section~\ref{sec:SM_compact_modes} illustrates compact Majorana zero modes bound to even-face defects and their dependence on the adjacent triangle fluxes. Section~\ref{sec:SM_chern_marker} describes the local Chern-marker calculation and the treatment of the degenerate zero-energy subspace.

\setcounter{tocdepth}{2}
\tableofcontents

\section{Ground-State Sector and Recursive Green's Function}
\label{sec:SM_ground}

\begin{figure}[t]
 \centering
 \includegraphics[width=0.96\linewidth]{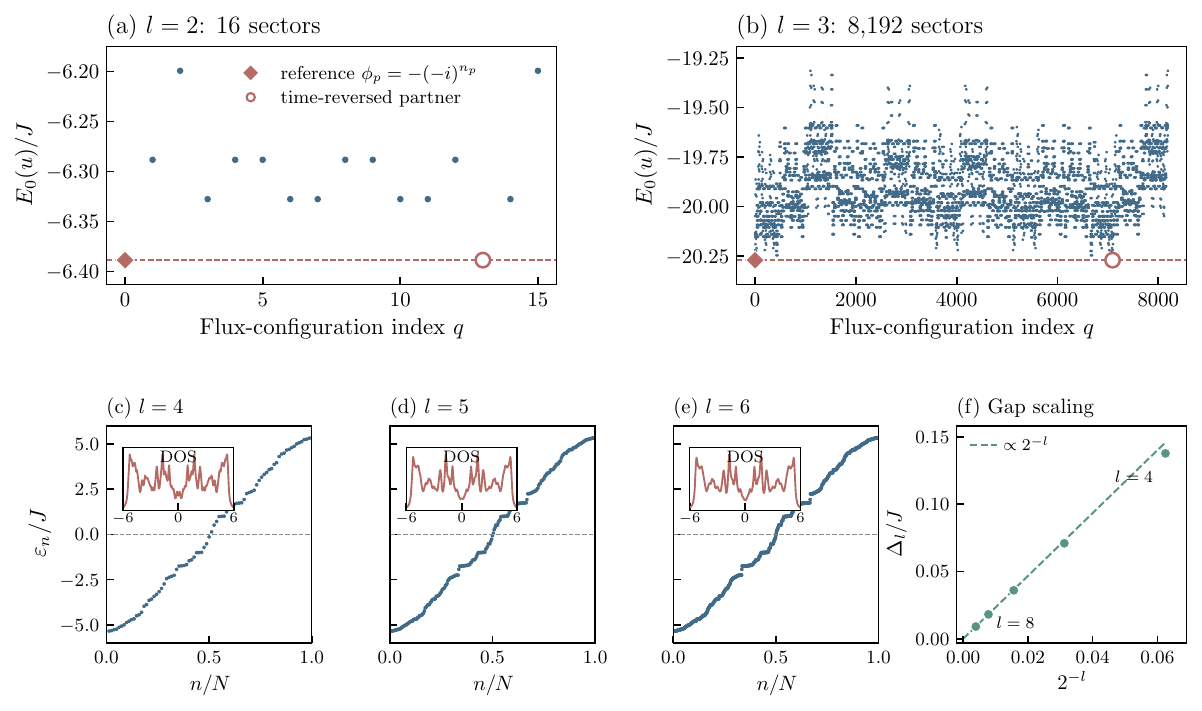}
 \caption{Testing the reference fluxes and their matter spectrum. (a,b) Exhaustive ground-energy comparison on the open $l=2,3$ gaskets, retaining all triangle and even-face flux choices. Each blue point is one gauge-inequivalent flux configuration; $q$ encodes its face-flip bits as defined in the text. The filled diamond marks the proposed reference and the open circle its time-reversed partner, the only two minima. Dashed lines mark their common energy, computed using Eq.~\eqref{eq:SM_sector_energy}. (c--e) Ordered-sector single-particle spectra with the auxiliary corner closure, for $l=4,5,6$. Here the horizontal coordinate is the normalized eigenvalue index $n/N$, with $N=3^l+1$ in this figure. Insets show the DOS per site, $N^{-1}\sum_n \delta_\eta(\omega-\varepsilon_n)$, using Lorentzian broadening $\eta/J=0.12$ and the same horizontal energy interval $\omega/J\in[-6,6]$. (f) Finite-size matter gap $\Delta_l=\min_n|\varepsilon_n|$ in the same auxiliary-closed reference sector, for $l=4,\ldots,8$, versus the inverse linear scale $2^{-l}$. The dashed line shows a $2^{-l}$ scaling.}
 \label{fig:SM_Ground_State}
\end{figure}

We use the Pauli-spin Hamiltonian of the main text,
\begin{equation}
 H=J\sum_{\langle ij\rangle_\gamma}\sigma_i^\gamma\sigma_j^\gamma
   =-iJ\sum_{\langle ij\rangle_\gamma}\hat u_{ij}c_ic_j,
 \qquad \hat u_{ij}=ib_i^\gamma b_j^\gamma,
 \label{eq:SM_spin_hamiltonian}
\end{equation}
where $\sigma^\gamma$ are Pauli matrices with eigenvalues $\pm1$, $\{c_i,c_j\}=2\delta_{ij}$, and each bond is counted once with a chosen orientation. Replacing the conserved link operators by their eigenvalues gives
\begin{equation}
 H_u=\frac{i}{4}\sum_{ij}A_{ij}(u)c_ic_j,
 \qquad A_{ij}(u)=-2Ju_{ij}=-A_{ji}(u),
 \qquad h(u)=iA(u).
 \label{eq:SM_single_particle_convention}
\end{equation}
The expression for $A_{ij}$ applies to an oriented bond, with zero entries elsewhere. While $H_u$ acts on the many-body matter state, the Hermitian matrix $h(u)$ determines the single-particle energies $\varepsilon_n$ and propagator $e^{A(u)t}=e^{-ih(u)t}$. We express energies in units of $J$ and times as $Jt$, with $\hbar=1$.

The reference sectors have the ordered fluxes specified in the main text, $\phi_p^{\rm gs}=-(\pm i)^{n_p}$. In the numerical plots below we traverse plaquettes counterclockwise and choose $\phi_p^{\rm gs}=-(-i)^{n_p}$, in particular $\phi_\triangle=-i$. The main text's clockwise traversal conjugates the triangle flux of the same link field, simply exchanging the two chirality labels; the even-face fluxes and the energy comparison are unchanged. Each flux pattern specifies a gauge-equivalence class of link fields.

To test this reference sector, we enumerate all flux patterns on the open $l=2$ and $l=3$ gaskets, allowing both triangle chiralities and all even-face fluxes. The bounded faces form an independent cycle basis, so assigning their fluxes enumerates the gauge-inequivalent sectors. For each pattern, we construct a compatible link field $u$ and diagonalize $h(u)$.

The ground energy is obtained from the negative single-particle levels with the Majorana factor of one half:
\begin{equation}
 E_0(u)=-\frac12\sum_{\varepsilon_n(u)>0}\varepsilon_n(u)
       =\frac12\sum_{\varepsilon_n(u)<0}\varepsilon_n(u)
       =-\frac14\operatorname{Tr}|h(u)|.
 \label{eq:SM_sector_energy}
\end{equation}
The factor $1/2$ follows because each pair $\pm\varepsilon$ represents one complex fermion with Hamiltonian $\varepsilon(f^\dagger f-1/2)$. Zero modes contribute no energy. The open gasket has dangling gauge Majoranas, whose zero-energy occupations can satisfy the physical parity constraint without exciting a positive-energy matter mode \cite{Zschocke2015}. Direct spin-basis diagonalization at $l=2$ reproduces the full flux-resolved Majorana spectrum, including its multiplicities.

Figure~\ref{fig:SM_Ground_State}(a,b) shows that the reference and its time-reversed partner are the only energy minima on both enumerated gaskets. They differ by reversing all triangle fluxes while leaving the even faces unchanged. The horizontal coordinate is the enumeration index $q=\sum_{p=0}^{N_{\rm f}-1}2^p f_p$, where $f_p=0,1$ specifies $\phi_p(q)=(-1)^{f_p}\phi_p^{\rm gs}$ for a fixed ordering of the $N_{\rm f}$ bounded faces. This finite-size test supports the ground-state flux rule used in the main text. 

Within the reference sector, the low-energy spectrum reflects the gasket's hierarchy of boundaries. An open level-$l$ gasket contains $N=3^l$ sites. Since a real antisymmetric matrix of odd dimension has zero determinant, its single-particle spectrum necessarily contains at least one exact zero mode, irrespective of topology. To distinguish this algebraic zero from the closing of other excitation energies, the spectral check in Fig.~\ref{fig:SM_Ground_State}(c--e) adds one auxiliary site and connects it to the three corners with hopping magnitude $2J$. The original plaquette fluxes are left unchanged; the auxiliary links are fixed to $u_{ja}=+1$ when oriented from corner $j$ to auxiliary site $a$. The resulting matrix has $3^l+1$ sites and no odd-dimension requirement for a zero eigenvalue. This auxiliary closure is used only for this spectral comparison; the spin correlations and wavepacket dynamics use the open gasket. The low-energy spacing decreases with generation, consistently with chiral modes on increasingly long boundaries. Panel (f) follows the smallest excitation energy $\Delta_l=\min_n|\varepsilon_n|$ through $l=8$. The gap decreases approximately as $2^{-l}$, consistent with the level spacing of chiral modes on boundaries whose lengths double at each generation. This scaling supports a gapless matter spectrum as $l\to\infty$; together with the boundary-mode dynamics and Chern marker in the main text, it is consistent with a gapless chiral spin liquid.

\begin{figure}[t]
 \centering
 \includegraphics[width=\linewidth]{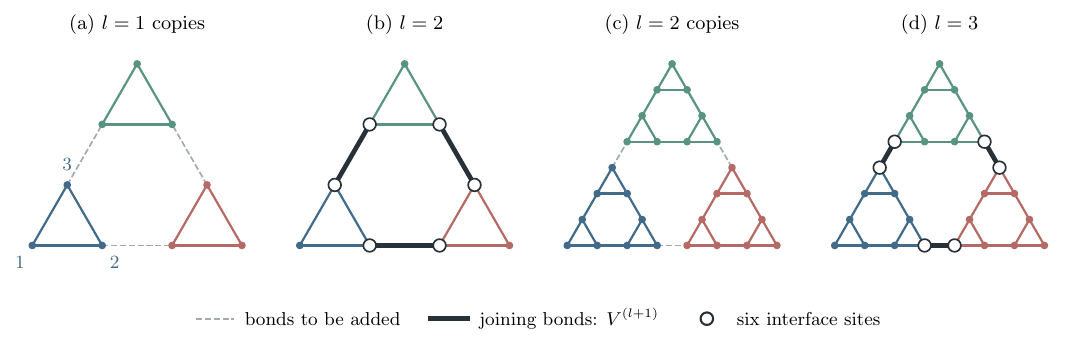}
 \caption{Recursive construction of the open KSG. Three elementary $l=1$ triangles before and after joining to form the $l=2$ gasket; sites 1--3 label the left triangle. (c) Three translated level-two gaskets; colors distinguish copies, not Kitaev bond types. Dashed links indicate the bonds still to be added. (d) The three joining bonds (thick dark lines) produce the level-three gasket and its new central twelve-sided face. Rings identify the six sites on which $V^{(3)}$ acts. The same joining step applies at every generation.}
 \label{fig:SM_recursion}
\end{figure}

The same geometric hierarchy provides a recursive construction of the Green's function. The open lattice is built by joining three disjoint level-$l$ gaskets with three new corner-to-corner bonds, as illustrated in Fig.~\ref{fig:SM_recursion}. Each copy retains its internal connectivity and flux pattern. Cyclic permutations of the bond colors make the joining colors compatible with the missing color at each corner; at isotropic $J$ these permutations do not change the single-particle hopping magnitudes. The new bonds enclose one additional central face of length $6\,2^{l-1}$. Choosing their signs to impose its reference flux preserves the inherited fluxes and constructs a representative of the ordered sector at the next level. With inherited site ordering and gauge on the three copies,
\begin{equation}
 h^{(l+1)}=h_0^{(l+1)}+V^{(l+1)},
 \qquad h_0^{(l+1)}=h^{(l)}\oplus h^{(l)}\oplus h^{(l)}.
 \label{eq:SM_recursive_h}
\end{equation}
Each joining bond contributes a Hermitian pair of matrix entries, $V_{ij}=-2iJu_{ij}$ and $V_{ji}=V_{ij}^*$. Thus $V$ couples only the six interface sites.

Define the single-particle Green function $\mathcal G^{(l)}(z)=[z-h^{(l)}]^{-1}$, where $z=\omega+i\eta$ and $\eta>0$. Here $\mathcal G$ denotes a Green function, whereas $G_{kL}^{(u)}$ in Sec.~\ref{sec:SM_spin} denotes an initial Majorana covariance. For the counterclockwise triangle $(1,2,3)$, choose $u_{12}=u_{23}=u_{31}=-1$, which gives $\phi_\triangle=-i$. Writing the hopping magnitude as $t_0=2J$, direct inversion gives
\begin{equation}
 h^{(1)}=t_0\begin{pmatrix}0&i&-i\\-i&0&i\\i&-i&0\end{pmatrix},
 \qquad
 \mathcal G^{(1)}(z)=\frac{1}{z(z^2-3t_0^2)}
 \begin{pmatrix}
 z^2-t_0^2&-t_0^2+it_0z&-t_0^2-it_0z\\
 -t_0^2-it_0z&z^2-t_0^2&-t_0^2+it_0z\\
 -t_0^2+it_0z&-t_0^2-it_0z&z^2-t_0^2
 \end{pmatrix}.
 \label{eq:SM_triangle_resolvent}
\end{equation}
Its poles are $0,\pm2\sqrt3J$. Reversing all triangle links gives the opposite-chirality Hamiltonian; a local gauge change conjugates the Green function by a diagonal sign matrix. The diagonal spectral weights are invariant under this change of gauge.

\begin{figure}[t]
 \centering
 \includegraphics[width=\linewidth]{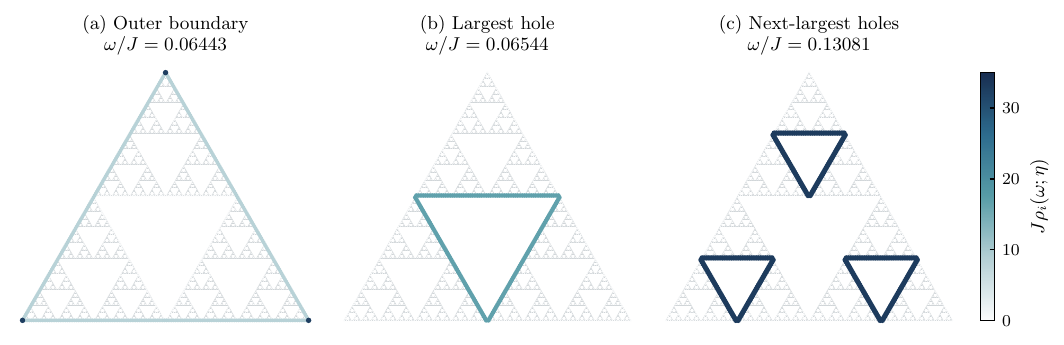}
 \caption{Spatially resolved LDOS in the ground-state flux sector of an open level-seven KSG. (a) Outer-boundary resonance. (b) Resonance on the largest internal hole. (c) Resonances on the three next-largest holes. The frequencies are indicated above the panels; each map evaluates Eq.~\eqref{eq:SM_LDOS} with Lorentzian width $\eta/J=10^{-4}$, retaining the full spectrum, including zero modes. Color gives $J\rho_i(\omega;\eta)$ on the same linear scale in all panels. Thin gray bonds show the underlying lattice.}
 \label{fig:SM_LDOS_maps}
\end{figure}

For the disconnected copies, $\mathcal G_0^{(l+1)}=\mathcal G^{(l)}\oplus \mathcal G^{(l)}\oplus\mathcal G^{(l)}$. Substituting Eq.~\eqref{eq:SM_recursive_h} into the Green function yields Dyson's identity and its solution \cite{Sivan1988},
\begin{equation}
 \mathcal G^{(l+1)}=\mathcal G_0^{(l+1)}+
 \mathcal G_0^{(l+1)}V^{(l+1)}\mathcal G^{(l+1)}
 =[1-\mathcal G_0^{(l+1)}V^{(l+1)}]^{-1}
 \mathcal G_0^{(l+1)}.
 \label{eq:Dyson}
\end{equation}
The six-site support of $V$ reduces the matrix inversion to the interface. Let $B$ be the matrix whose columns are the six interface-site basis vectors and set $V_b=B^\dagger V B$, so that $V=BV_bB^\dagger$. Suppressing the generation label, one obtains
\begin{equation}
 \mathcal G=\mathcal G_0+
 \mathcal G_0 B[1-V_b B^\dagger\mathcal G_0 B]^{-1}
 V_b B^\dagger\mathcal G_0.
 \label{eq:SM_interface_dyson}
\end{equation}
The local density of states then follows from
\begin{equation}
 \rho_i(\omega;\eta)=-\frac{1}{\pi}\operatorname{Im}
 \mathcal G_{ii}(\omega+i\eta)
 \label{eq:SM_LDOS}
\end{equation}
with normalized eigenvectors $\varphi_n$. This gives the local spectral weights, while $N^{-1}\sum_i\rho_i$ gives the DOS per site. This recursion applies to the open gasket and agrees with direct matrix inversion on the sizes used for numerical checks.
The LDOS maps in Fig.~\ref{fig:SM_LDOS_maps} locate the outer and inner chiral modes by LDOS defined in Eq.~\ref{eq:SM_LDOS}. They evaluate $\rho_i(\omega;\eta)$, defined in Eq.~\eqref{eq:SM_LDOS} below. At representative low-energy resonances of the level-seven ground-state sector, the spectral weight follows the outer perimeter, the largest internal hole, or the next-largest holes. These profiles exhibit the hierarchy of boundary modes on the gasket, with spectral weight confined transversely but spread around a selected boundary scale. Their spatial support complements the energy spectrum; the propagation direction is revealed by the dynamics discussed below.

\section{Extended and Localized Modes: Inverse Participation Ratios}
\label{sec:SM_ipr}

To distinguish localized matter states from modes that extend along the gasket boundaries, we diagonalize the open KSG in three ensembles: the ground-state flux sector, independent even-face flux flips with probability $1/2$ at fixed triangle chirality, and independent flips of all faces with probability $1/2$. For a normalized eigenmode of $h(u)=iA(u)$, define
\begin{equation}
 P_{2,n}^{(u)}=\sum_{j=1}^{N}|\varphi_{n,j}^{(u)}|^4,
 \qquad \frac{1}{N}\leq P_{2,n}^{(u)}\leq 1,
 \qquad N=3^l.
 \label{eq:SM_ipr_definition}
\end{equation}
A state confined to a fixed number of sites has a nonvanishing IPR as $l$ grows; a state uniformly spread over the whole gasket has $P_2\sim N^{-1}$. A boundary mode is different. The outer perimeter contains $L_\partial=3(2^l-1)$ sites, so a mode extended along that perimeter but confined transversely satisfies
\begin{equation}
 P_2\sim L_\partial^{-1}\sim 2^{-l}
       \sim N^{-\ln2/\ln3},
 \label{eq:SM_boundary_ipr}
\end{equation}
rather than $N^{-1}$. Modes circulating around a hole of fixed size can have finite IPR even though they propagate along that hole. Thus IPR must be interpreted together with the spatial support, especially on a lattice with a hierarchy of internal boundaries \cite{Pai2019}.

At fixed $u$, this Hamiltonian belongs to symmetry class D \cite{AltlandZirnbauer1997,EversRMP2008}. Since $A$ is real and antisymmetric, complex conjugation gives $h(u)^*=-h(u)$, with particle-hole operation squaring to $+1$. Time reversal changes the triangle fluxes and is not a symmetry within a generic fixed sector. The triangles also preclude bipartite sublattice symmetry. These properties hold for each realization, including in the all-face ensemble with statistically balanced triangle chiralities. Localization depends on both symmetry and geometry \cite{EversRMP2008}; here we assess it from finite-size scaling and spatial support.

\begin{figure}[t]
 \centering
 \includegraphics[width=\linewidth]{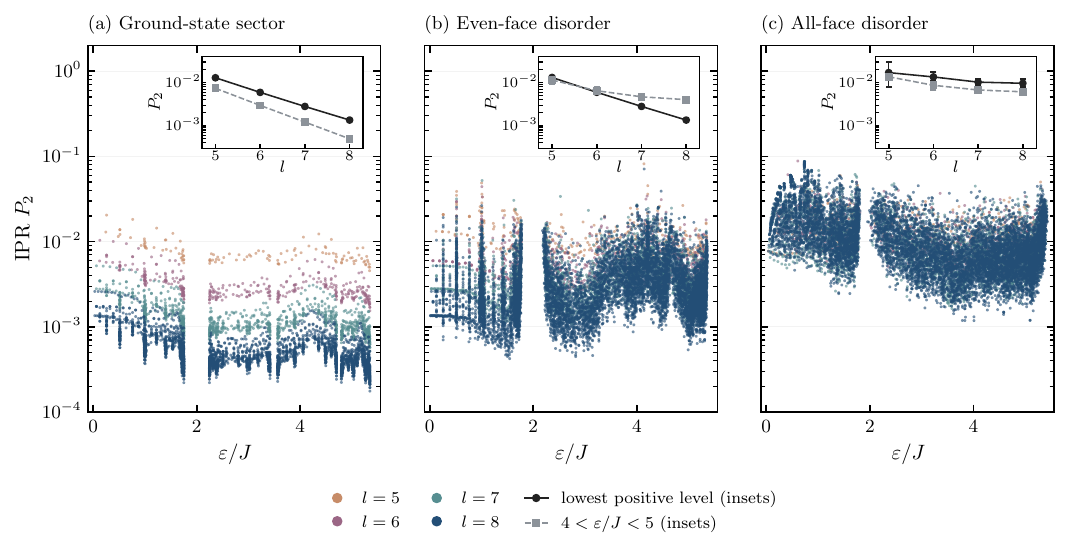}
 \caption{Energy-resolved IPR on open KSGs with $l=5,6,7,8$ ($N=243,729,2187,6561$). (a) Ground-state flux sector. (b) Even-face flux disorder, with uniform triangle chirality. (c) All-face flux disorder. Each eligible face is flipped independently with probability $1/2$, using three realizations per disordered ensemble and size. Dots show individual positive-energy eigenstates; their negative-energy partners have identical IPR. Zero-energy states are omitted because degeneracy makes the IPR of individual zero-mode eigenvectors basis dependent. Insets compare the IPR of the lowest positive-energy state (circles) with the geometric-mean IPR in $4<\varepsilon/J<5$ (squares). Points are realization means and whiskers span the observed range; the clean sector has only one value. Exact zero modes and numerically degenerate nonzero levels are excluded from the high-energy statistic. All energies use the Pauli-spin exchange convention of Sec.~\ref{sec:SM_ground}.}
 \label{fig:SM_ipr}
\end{figure}

\begin{figure}[t]
 \centering
 \includegraphics[width=0.96\linewidth]{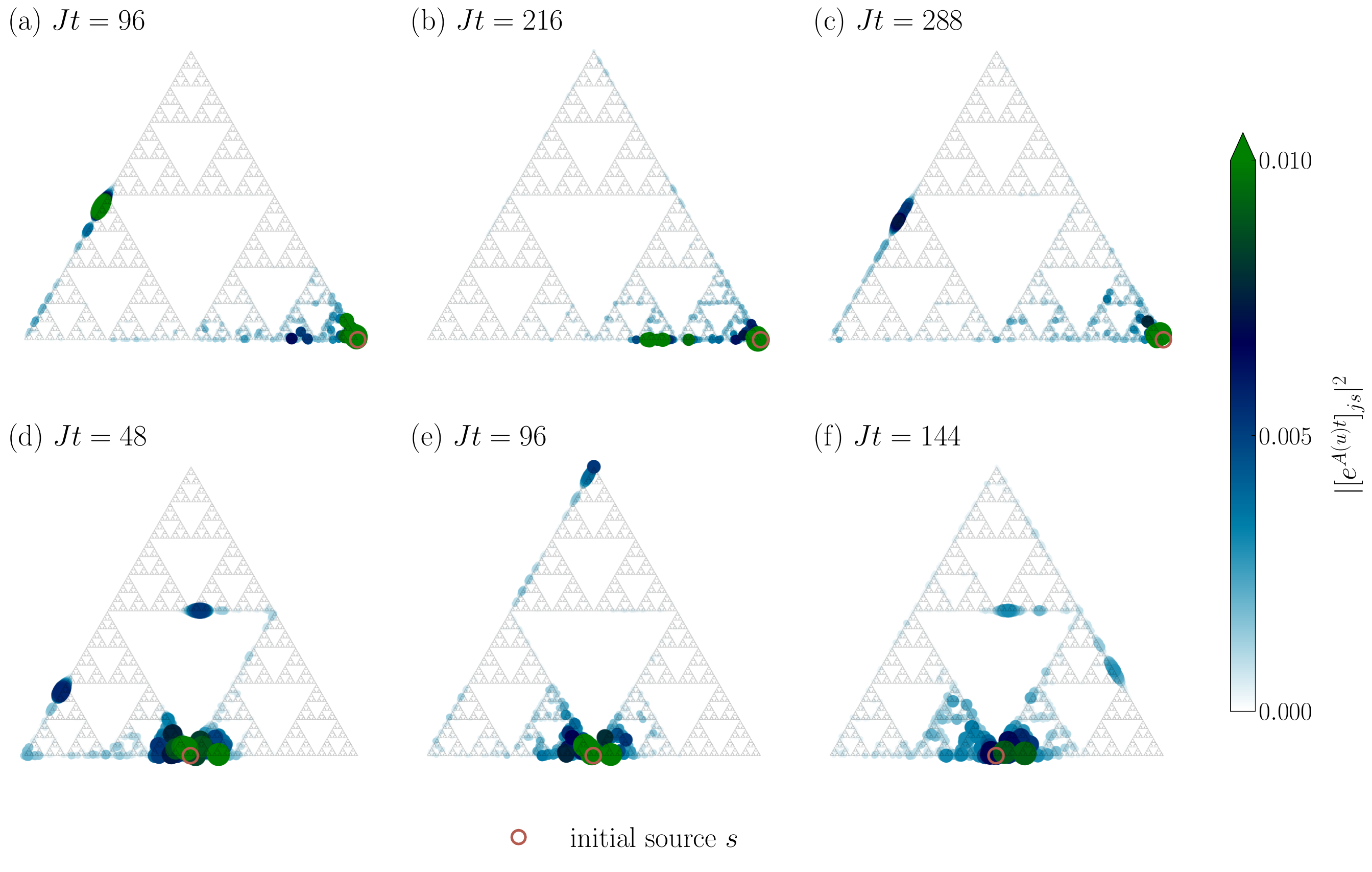}
 \caption{Single-particle matter propagation in one even-flux-disordered level-eight KSG ($N=6561$), showcasing the coexistence of localized modes and chiral edge modes. (a--c) A delta source at the right outer corner. (d--f) A delta source at the lower midpoint of the largest internal hole (the left endpoint of its bottommost bond). Open red rings mark the source in each row. Color shows the absolute probability $|[e^{A(u)t}]_{js}|^2$ on a common linear scale, saturated above $0.01$. Marker area is proportional to probability divided by its maximum within each snapshot. Times are given in the Pauli-spin convention $Jt$. }
 \label{fig:dynamics}
\end{figure}

Figure~\ref{fig:SM_ipr} compares the nonzero-energy IPR through $l=8$. The insets separate the lowest positive-energy level from a fixed high-energy window; neither statistic includes the exact zero modes. We also compute the probability on the outer perimeter and on the boundary of the largest hole, rather than identifying an edge state from its energy alone. The ground-state sector provides a reference for the size dependence in the absence of flux disorder.
In the ground-state sector and under even-face disorder, the lowest positive-energy state follows approximately inverse-perimeter scaling in Eq.~\eqref{eq:SM_ipr_definition}. Its probability remains concentrated on the outer boundary as the gasket grows. Thus even-face disorder preserves a low-energy mode extended along that boundary.

At higher energies, the typical IPR decreases much more slowly with size in both disordered ensembles than in the clean sector. This behavior indicates a stronger tendency toward localization, in contrast to the continuing boundary scaling of the low-energy mode under even-face disorder. Under all-face disorder, the lowest levels also lose the clean sector's strong outer-boundary concentration and the states tend to localize across the whole spectrum. The accessible sizes therefore reveal a distinction between the two kinds of flux disorder, and a strong fractal-induced localization effect across the whole spectrum compared to its two-dimensional counterparts.

\section{Single-Particle Time Evolution}
\label{sec:SM_wavepacket}

To visualize what can propagate in a fixed flux background, we initialize a single-particle basis vector on a site $s$ and compute
\begin{equation}
 \psi_j(t;u)=[e^{-ih(u)t}]_{js}=[e^{A(u)t}]_{js},
 \qquad \psi_j(0;u)=\delta_{js},
 \qquad \sum_j|\psi_j(t;u)|^2=1.
 \label{eq:SM_packet}
\end{equation}
The intensity $|\psi_j(t;u)|^2$ diagnoses propagation in the quadratic matter Hamiltonian. It is gauge invariant: under $c_j\mapsto s_jc_j$, with $s_j=\pm1$, the propagator acquires only the endpoint signs $s_js_s$. A physical spin response additionally depends on the initial spin preparation, its Majorana covariance, and its flux-sector probabilities. Section~\ref{sec:SM_spin} derives how these ingredients enter $C_{LV}(t)$.

Figure~\ref{fig:dynamics} shows an open level-eight gasket in one representative sector: every triangle has $\phi_\triangle=-i$, and each even-face flux is independently flipped from the ordered sector with probability $1/2$. The two rows use the same Hamiltonian but different source sites, at the right outer corner and at the lower midpoint of the largest internal hole. The latter midpoint is a short horizontal bond; we choose its left endpoint. A site-localized source overlaps the full spectrum, producing both circulating boundary weight and a residual contribution near its origin. This coexistence agrees with the participation-ratio and boundary-weight analysis of Sec.~\ref{sec:SM_ipr}. The snapshots show propagation in a single realization, before the flux-sector average required for the spin response.

\section{Gauge-Invariant Dynamics of Corner Spin Correlations}
\label{sec:SM_spin}
\label{sec:sector_decomp}
\label{sec:SM_endpoint}
\label{sec:prop_formula}
The wavepacket analysis isolates matter propagation in a given flux background. To obtain the physical spin response, we must also include the initial state and its distribution over those backgrounds. Here we derive the exact corner-correlation formula used in the main text. The essential ingredients are the flux-preserving corner probes, the prepared state's definite $P_z$, and the path connecting the dangling Majoranas on the gasket. We then derive the sector probabilities for $|\Psi_{1,2}\rangle$ and their partially flux-resolved descendants $|\Psi_3\rangle$.

Let $|\Psi\rangle$ be the normalized initial spin state and $|\widetilde\Psi\rangle$ its physical Majorana embedding, including both gauge and matter degrees of freedom. Physicality means $D_j|\widetilde\Psi\rangle=|\widetilde\Psi\rangle$, where $D_j=b_j^xb_j^yb_j^zc_j$ \cite{Kitaev2006,Zschocke2015}. The conserved physical links are resolved by the commuting projectors
\begin{equation}
 \Pi_u=\prod_{\ell\ {\rm physical\ links}}
       \frac{1+u_\ell\hat u_\ell}{2},
 \qquad \sum_u\Pi_u=1,
 \qquad \Pi_u\Pi_{u'}=\delta_{uu'}\Pi_u.
 \label{eq:SM_link_projector}
\end{equation}
For sectors with nonzero weight, define
\begin{equation}
 p(u)=\langle\widetilde\Psi|\Pi_u|\widetilde\Psi\rangle,
 \qquad |\widetilde\Psi_u\rangle
       =\frac{\Pi_u|\widetilde\Psi\rangle}{\sqrt{p(u)}},
 \qquad |\widetilde\Psi\rangle
       =\sum_u\sqrt{p(u)}\,|\widetilde\Psi_u\rangle.
 \label{eq:sector_decomp}
\end{equation}
The normalized vector $|\widetilde\Psi_u\rangle$ retains all unresolved matter and dangling-Majorana degrees of freedom. Physicality constrains the full embedded state; a local $D_j$ maps an individual link component to the component with the adjacent link signs reversed.

With $L$ the insertion corner and $V$ the observation corner, the correlator is
\begin{equation}
 C_{LV}(t)=\langle\Psi|\sigma_V^{\gamma_V}(t)
                         \sigma_L^{\gamma_L}|\Psi\rangle,
 \qquad \sigma_j^{\gamma_j}=i b_j^{\gamma_j}c_j.
 \label{eq:CRL_def_SM}
\end{equation}
We choose $\gamma_j$ to be the missing bond color at each probed corner. The corresponding $b_j^{\gamma_j}$ is absent from the Hamiltonian and is conserved. The spin $i b_j^{\gamma_j}c_j$ commutes with every physical link: its two Majoranas contribute canceling signs on the present links. Since the Hamiltonian also conserves the links,
\begin{equation}
 [\sigma_L^{\gamma_L},\Pi_u]
 =[\sigma_V^{\gamma_V}(t),\Pi_u]=0.
 \label{eq:sigma_sector_diag_SM}
\end{equation}
Thus each spin insertion preserves the flux background, and orthogonality of the link sectors gives
\begin{align}
 C_{LV}(t)
 =\sum_u p(u)\langle\widetilde\Psi_u|
        \sigma_V^{\gamma_V}(t)\sigma_L^{\gamma_L}
                     |\widetilde\Psi_u\rangle 
 =\sum_u\langle\widetilde\Psi|\Pi_u
        \sigma_V^{\gamma_V}(t)\sigma_L^{\gamma_L}
                     |\widetilde\Psi\rangle.
 \label{eq:CRL_incoherent_exact_SM}
\end{align}
This probability-weighted sum is exact for the coherent initial spin state. Dynamical dephasing subsequently arises from differences between the sector-resolved amplitudes.

Within a gauge sector, the spin product becomes
\begin{equation}
 \Pi_u\sigma_V^{\gamma_V}(t)\sigma_L^{\gamma_L}\Pi_u
 =\Pi_u b_V^{\gamma_V}b_L^{\gamma_L}c_V(t;u)c_L\Pi_u,
 \qquad c_V(t;u)=e^{iH_ut}c_Ve^{-iH_ut}.
 \label{eq:sigma_to_c_SM}
\end{equation}
The positive sign follows from $i^2=-1$ and the anticommutation of $b_L^{\gamma_L}$ with $c_V(t;u)$. This anticommutation holds at all times because quadratic evolution expresses $c_V(t;u)$ as a linear combination of matter Majoranas. For any sector-preserving operator $O$, projector idempotence gives
\begin{equation}
 \langle\widetilde\Psi_u|O|\widetilde\Psi_u\rangle
 =\frac{\langle\widetilde\Psi|\Pi_u O\Pi_u
                      |\widetilde\Psi\rangle}{p(u)}
 =\frac{\langle\widetilde\Psi|\Pi_u O
                      |\widetilde\Psi\rangle}{p(u)}.
 \label{eq:SM_projector_equivalence}
\end{equation}
In particular, $c_kc_L$ commutes with every link, so the initial Majorana covariance is
\begin{equation}
 G_{kL}^{(u)}
 =\langle\widetilde\Psi_u|c_kc_L|\widetilde\Psi_u\rangle
 =\frac{\langle\widetilde\Psi|\Pi_u c_kc_L
                      |\widetilde\Psi\rangle}{p(u)}.
 \label{eq:SM_G_projector}
\end{equation}
Both $G_{kL}^{(u)}$ and $p(u)$ are therefore fixed by the physical spin preparation.

For $V=L$, the dangling pair in Eq.~\eqref{eq:sigma_to_c_SM} squares to one. For distinct corners, the gasket geometry and spin parity determine its action. Each triangle state in $|\Psi_1\rangle$ and $|\Psi_2\rangle$ contains only zero or two down spins, giving $P_{z,\triangle}=\prod_{j\in\triangle}\sigma_j^z=+1$ and hence $P_z=\prod_j\sigma_j^z=+1$ for the full state. This parity is also preserved when preparing $|\Psi_3\rangle$: conjugation by $P_z$ changes the sign of both spins on an $x$ or $y$ bond and neither on a $z$ bond. Every plaquette operator, and therefore every flux-measurement projector, commutes with $P_z$.

Take $L$ and $V$ to be the corners with missing colors $y$ and $x$, respectively. The $x/y$ bonds form a single path through every site of the gasket, with precisely these two dangling endpoints (End Matter B). Using $\sigma_j^zD_j=-ib_j^xb_j^y$ on the physical state and pairing the internal Majoranas along the path then gives
\begin{equation}
 q_z|\widetilde\Psi\rangle
 =\prod_j(-ib_j^xb_j^y)|\widetilde\Psi\rangle
 =\eta i\,b_V^x b_L^y
     \prod_{\ell\in\{x,y\}}\hat u_\ell
       |\widetilde\Psi\rangle,
 \label{eq:SM_Pz_identity}
\end{equation}
where $P_z|\widetilde\Psi\rangle=q_z|\widetilde\Psi\rangle$ and $\eta=\pm1$ is fixed by site ordering and link orientations. Applying $\Pi_u$ replaces each $\hat u_\ell$ by $u_\ell$, yielding
\begin{equation}
 b_V^x b_L^y|\widetilde\Psi_u\rangle
 =q_z\chi_{LV}^{(u)}|\widetilde\Psi_u\rangle,
 \qquad \chi_{LV}^{(u)}=-i\eta
                  \prod_{\ell\in\{x,y\}}u_\ell.
 \label{eq:SM_pair_reduction}
\end{equation}
The endpoint pair thus acts as a sector-dependent scalar determined jointly by the links and the prepared spin parity. Since it commutes with every matter bilinear, Eq.~\eqref{eq:SM_pair_reduction} removes it from the correlation matrix element. A pair involving the $z$-missing corner and either the $x$- or $y$-missing corner instead has a vanishing correlation: its spin product is odd under $P_z$. This accounts for the zero signals in the main text's spin-ED benchmark.

It remains to evolve the matter operator. From $[c_mc_n,c_j]=2(\delta_{nj}c_m-\delta_{mj}c_n)$ and the antisymmetry of $A$, Eq.~\eqref{eq:SM_single_particle_convention} gives
\begin{equation}
 [H_u,c_j]=-i\sum_k A_{jk}(u)c_k,
 \qquad \frac{d c_j(t;u)}{dt}=\sum_k A_{jk}(u)c_k(t;u),
 \qquad c_V(t;u)=\sum_k[e^{A(u)t}]_{Vk}c_k.
 \label{eq:majorana_eom_SM}
\end{equation}
Combining the sector sum, the endpoint reduction, and this propagator for the preparations with $q_z=+1$ gives the main text's exact formula,
\begin{equation}
 C_{LV}(t)=\sum_{u,k}p(u)\chi_{LV}^{(u)}
                      [e^{A(u)t}]_{Vk}G_{kL}^{(u)},
 \qquad \chi_{LL}^{(u)}=1.
 \label{eq:CRL_master_formula_SM}
\end{equation}
Here $V=L$ or $L,V$ are the distinct $y$- and $x$-missing corners. The reduction uses only the operator algebra and the prepared spin parity, and applies without a Gaussian-state assumption.

The complete expression is gauge invariant. Under site signs $s_j$, $G_{kL}^{(u)}$ acquires $s_ks_L$, the propagator acquires $s_Vs_k$, and the path factor acquires $s_Vs_L$; these signs cancel in each summand. One can therefore use a consistent representative of each physical flux sector, with its corresponding covariance and endpoint factor. For samples drawn from the preparation's Born distribution, the estimator and the plotted intensity are
\begin{equation}
 C_{LV}^{(N_s)}(t)=\frac{1}{N_s}\sum_{a=1}^{N_s}
 \chi_{LV}^{(u_a)}\sum_k[e^{A(u_a)t}]_{Vk}G_{kL}^{(u_a)},
 \qquad \big|C_{LV}^{(N_s)}(t)\big|^2.
 \label{eq:SM_sampling}
\end{equation}
Averaging amplitudes before taking the absolute square retains the interference between responses in different physical flux sectors. Their relative phases can suppress the cross-corner signal, while partial flux measurements narrow the contributing ensemble. The spin-ED comparison in the main text checks the full correlation, including the initial-state covariance and sector weights.

\section{Compact Majorana Zero Modes on Even-Face Defects}
\label{sec:SM_compact_modes}

\begin{figure}[H]
 \centering
 \includegraphics[width=0.65\linewidth]{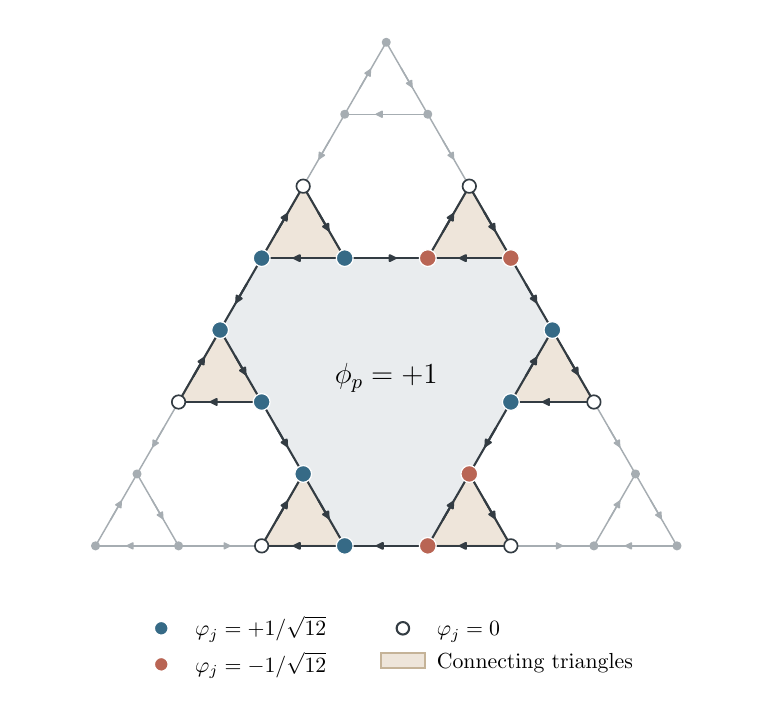}
 \caption{An exact compact zero mode on the largest hole of an $l=3$ gasket. The central label identifies the single excited flux $\phi_p=+1$; all unlabelled faces retain their reference fluxes. An arrow $j\to k$ denotes $u_{jk}=+1$, with $A_{jk}=-2Ju_{jk}$. Blue and rust circles give the signed wavefunction amplitudes shown in the legend. The shaded connecting triangles have zero-amplitude tips, marked by open circles; all remaining background sites also have zero amplitude.}
 \label{fig:SM_compact_construction}
\end{figure}

An even-face flux defect can bind a Majorana mode whose support ends exactly at the hole boundary. Figure~\ref{fig:SM_compact_construction} gives a concrete example on the open $l=3$ gasket: only the largest, twelve-sided hole is flipped from the reference sector, while all triangle fluxes retain their common chirality. In the displayed gauge, a normalized real wavefunction has amplitudes $\varphi_j=\pm1/\sqrt{12}$ on the twelve boundary sites, with the signs indicated by color, and vanishes on every other site. The corresponding Majorana operator is
\begin{equation}
 \gamma_p=\sum_j\varphi_jc_j,
 \qquad A(u)\varphi=0,
 \qquad \sum_j\varphi_j^2=1.
 \label{eq:SM_compact_operator}
\end{equation}
Thus $[H_u,\gamma_p]=0$ and $\gamma_p^2=1$. The arrows and amplitudes in the figure satisfy the zero-energy equation at every site of the full gasket, including sites outside the proposed support.

The six shaded triangles make the confinement transparent. Each has two sites on the supporting ring and one zero-amplitude tip connecting it to the rest of the gasket. At such a tip $i$, the two contributions cancel, $A_{ia}\varphi_a+A_{ib}\varphi_b=0$. The same amplitudes satisfy the equations along the ring. This destructive interference prevents leakage through every connection to the surrounding lattice. Reversing the flux of any one of these adjacent triangles changes the sign constraints and removes the twelve-site solution: the zero-mode equation restricted to these twelve sites then has only the zero solution. A triangle defect away from the supporting ring leaves this local solution unchanged. Consequently, unrestricted triangle disorder removes the uniform-chirality condition that protects the even-defect construction, although individual flux patterns can still support other localized zero modes.

\begin{figure}[t]
 \centering
 \includegraphics[width=0.9\linewidth]{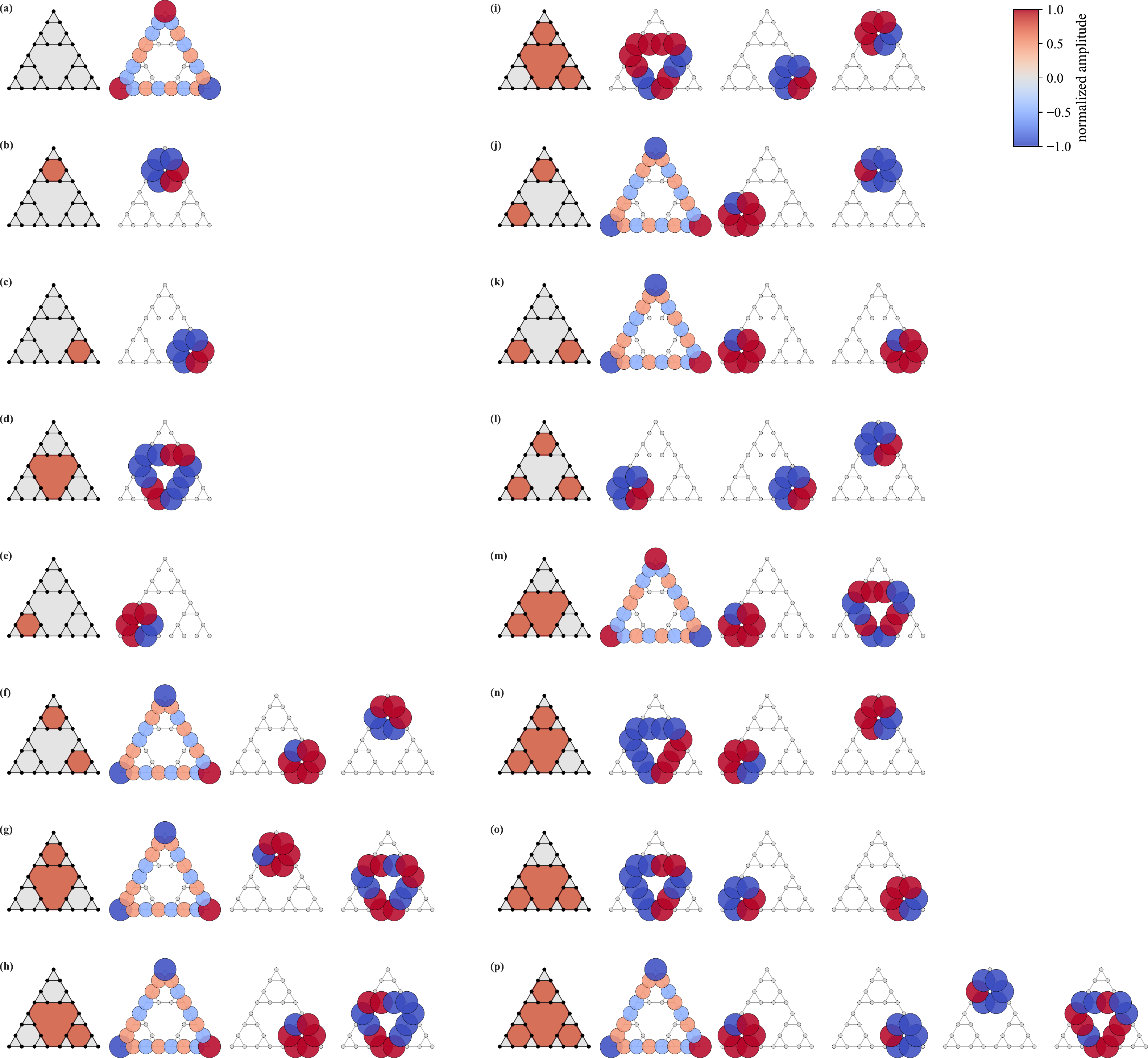}
 \caption{Complete enumeration of the sixteen even-face flux sectors on the open $l=3$ gasket, with uniform triangle chirality. Each labelled group starts with its flux configuration, followed by a complete zero-mode basis for that configuration. Colored hole interiors identify flipped even faces. Mode colors show signed amplitudes divided by the largest magnitude in each vector, and marker size increases with their magnitude. Note for even number of flipped even plaquettes, there is an additional zero mode supported on the whole boundary, which, however, does not hybridize with other zero modes either, thus does not change the results in the main text. }
 \label{fig:SM_compact_enumeration}
\end{figure}

Figure~\ref{fig:SM_compact_enumeration} collects complete zero-mode bases for all even-face flux assignments at $l=3$, including compact defect states and any additional gasket-wide mode allowed by the odd site number. Compact solutions survive simultaneous even-face flips. An eigensolver can mix a degenerate kernel into spatially broad vectors; the sparse basis makes its compact content visible. The general presence of compact MZM and its proof are developed in the companion work~\cite{companion_work}.

\clearpage

\section{Local Chern Marker and the Zero-Energy Subspace}
\label{sec:SM_chern_marker}

We characterize the matter topology using the real-space Chern marker of Bianco and Resta~\cite{Bianco11}, which follows the projector formulation of the Chern invariant~\cite{Kitaev2006}. For each link sector we diagonalize $h(u)=iA(u)$ and form the projector onto all strictly negative-energy states,
\begin{equation}
 P_-^{(u)}=\sum_{\varepsilon_n^{(u)}<0}
 |\varphi_n^{(u)}\rangle\langle\varphi_n^{(u)}|.
 \label{eq:SM_negative_projector}
\end{equation}
With $X_{ij}=x_i\delta_{ij}$ and $Y_{ij}=y_i\delta_{ij}$ the position operators of the planar embedding, the site-resolved quantity plotted in the main text is
\begin{equation}
 C_i[P_-]=4\pi\,\operatorname{Im}
 \langle i|P_-XP_-YP_-|i\rangle.
 \label{eq:SM_local_chern}
\end{equation} 
A local change of gauge conjugates $P_-$ by a diagonal sign matrix, which commutes with $X$ and $Y$; hence every $C_i$ is gauge invariant. The overall sign depends on the coordinate orientation and the chosen triangle chirality.

The trace over a finite open sample vanishes, since it is proportional to the trace of a commutator of projected position operators~\cite{Bianco11}. Boundary contributions therefore compensate the interior marker. For the estimates quoted in the main text, we omit a narrow outer-perimeter strip and divide the retained marker by its associated embedding area. Each polygonal face contributes its area equally to its boundary sites, giving $a_i=\sum_{p\ni i}\operatorname{Area}(p)/n_p$, and the reported estimate is $C=\sum_{i\in\mathcal B}C_i/\sum_{i\in\mathcal B}a_i$ over the retained sites $\mathcal B$. This geometric normalization makes the estimate dimensionless and independent of a uniform rescaling of the drawing. The ground-state and even-disordered sectors give values close to the same chiral reference value, whereas all-face disorder suppresses the interior average. These comparisons use the same geometry and averaging prescription.

Exact zero modes require a consistent occupation convention. We separate the numerically zero eigenvalues and exclude their entire eigenspace when constructing $P_-$. Selecting an arbitrary fraction of a degenerate zero-mode basis would make the result depend on the eigenvectors returned by diagonalization. The two complete-subspace choices, $P_-$ and $P_-+P_0$, are basis independent, with $P_0$ the zero-mode projector. They also give identical local markers here. Since $h^*=-h$, the positive-energy projector is $P_+=P_-^*$ and $P_-+P_0=1-P_+$. For real position operators, complex conjugation and complementation each reverse Eq.~\eqref{eq:SM_local_chern}, so
\begin{equation}
 C_i[P_-+P_0]=C_i[1-P_+]=-C_i[P_+]=C_i[P_-].
 \label{eq:SM_zero_marker_equivalence}
\end{equation}
This equality holds pointwise and includes the odd-site zero mode as well as defect-bound modes. The compact states can therefore generate a large zero-energy DOS without introducing an ambiguity into the marker. Their finite support prevents an extended zero-energy band from forming under even-face disorder, consistently with the persistence of the chiral marker in the main text.

\bibliography{Fractals}